\RequirePackage{fix-cm} % Fix LaTeX2e bugs.
\documentclass[a4paper, twoside, reqno, dvips, 12pt]{amsart}
\usepackage{fixltx2e}   % Fix LaTeX2e bugs.

\usepackage{etex}

\usepackage[latin1]{inputenc}
\usepackage[T1]{fontenc}

\usepackage{eucal}
\usepackage{esint}
\usepackage{dsfont}
\usepackage{xspace}
\usepackage{amsgen}
\usepackage{amsthm}
\usepackage{amssymb}
\usepackage{amsmath}
\usepackage{upgreek}
\usepackage{wasysym}
\usepackage{amsfonts}
\usepackage{stmaryrd}
\usepackage{mathtools}

\usepackage{mathrsfs}
\DeclareMathAlphabet{\mathscrbf}{OMS}{mdugm}{b}{n}

\usepackage{a4wide}

\usepackage[dvipsnames, table]{xcolor}
\definecolor{bckg}{RGB}{20.8, 20.8, 20.8}
\definecolor{oneblue}{rgb}{0.0, 0.0, 0.85}
\definecolor{Lightblue}{RGB}{214, 214, 214}
\definecolor{bluepigment}{rgb}{0.2, 0.2, 0.6}
\definecolor{charcoal}{rgb}{0.21, 0.27, 0.31}
\definecolor{denimblue}{rgb}{0.08, 0.38, 0.74}
\definecolor{Lightgray}{rgb}{0.89, 0.89, 0.89}
\definecolor{darkgrey}{rgb}{0.273, 0.281, 0.30}
\definecolor{darkelectricblue}{rgb}{0.33, 0.41, 0.47}

\usepackage{psfrag}
\usepackage{graphicx}
\usepackage{subfigure}
\usepackage{morefloats}
\usepackage{indentfirst}

\usepackage{acronym}
\usepackage{microtype}
\usepackage[labelsep=period,%
            labelfont={bf,sf,color=bluepigment},%
            justification=raggedright]{caption}

\usepackage[perpage, symbol]{footmisc}

\usepackage[usenames, dvipsnames]{pstricks}
\usepackage{epsfig}
\usepackage{pst-grad} % For gradients
\usepackage{pst-plot} % For axes

\usepackage[colorlinks,
           urlcolor=oneblue,
           linkcolor=denimblue,
           citecolor=NavyBlue,
           bookmarksopen=false,
           pdfpagemode=UseNone,
           pagebackref]{hyperref}

\usepackage[sort&compress, comma, square, numbers]{natbib}

\usepackage[explicit]{titlesec}

\titleformat{\section}
  {\color{NavyBlue}\Large\sffamily\bfseries}
  {}
  {0em}
  {\colorbox{bckg!5}{\parbox{\dimexpr\linewidth-2\fboxsep\relax}{\centering\thesection. #1}}}
  [\vspace*{0.33em}]

\titleformat{name=\section,numberless}
  {\color{NavyBlue}\Large\sffamily\bfseries}
  {}
  {0.0em}
  {\colorbox{bckg!10}{\parbox{\dimexpr\linewidth-2\fboxsep\relax}{\centering#1}}}
  [\vspace*{0.33em}]

\titleformat{\subsection}
  {\color{NavyBlue}\large\sffamily\bfseries}
  {}
  {0.0em}
  {\colorbox{bckg!5}{\parbox{\dimexpr\linewidth-2\fboxsep\relax}{\centering\thesubsection. #1}}}
  [\vspace*{0.33em}]

\titleformat{name=\subsection,numberless}
  {\color{NavyBlue}\Large\sffamily\bfseries}
  {}
  {0em}
  {\colorbox{bckg!10}{\parbox{\dimexpr\linewidth-2\fboxsep\relax}{\centering#1}}}
  [\vspace*{0.33em}]

\titleformat{\subsubsection}
  {\color{bluepigment}\sffamily\normalsize\bfseries}
  {\thesubsubsection}
  {0.5em}
  {#1}
  [\vspace*{0.33em}]

\titleformat{\paragraph}[runin]
  {\color{bluepigment}\sffamily\small\bfseries}
  {}
  {0em}
  {#1}

\titlespacing{\section}{1.0em}{1.5em plus 2pt minus 2pt}%
{1.0em plus 2pt minus 2pt}[0em]
\titlespacing{\subsection}{1.0em}{1.5em plus 2pt minus 2pt}%
{1.0em}[0em]
\titlespacing{\subsubsection}{1.0em}{1.5em plus 2pt minus 2pt}%
{1.0em plus 2pt minus 2pt}[0em]

\usepackage{titletoc}

\contentsmargin{0.5em}
\newlength{\tocsep} 
\titlecontents{section}[\tocsep]
  {\addvspace{10pt}\bfseries\sffamily}
  {\contentslabel[\thecontentslabel]{\tocsep}}
  {}
  {\ \titlerule*[0.75pc]{.}\ \thecontentspage}
  []
\titlecontents{subsection}[\tocsep]
  {\addvspace{8pt}\sffamily}
  {\contentslabel[\thecontentslabel]{\tocsep}}
  {}
  {\ \titlerule*[0.5pc]{.}\ \thecontentspage}
  []
\titlecontents*{subsubsection}[\tocsep]
  {\addvspace{2pt}\footnotesize\sffamily}
  {}
  {}
  {\ \titlerule*[0.35pc]{.}\ \thecontentspage}
  [\\*]

\makeatletter
\def\@setauthors{%
  \begingroup
  \def\thanks{\protect\thanks@warning}%
  \trivlist
  \centering\footnotesize \@topsep30\p@\relax
  \advance\@topsep by -\baselineskip
  \item\relax
  \author@andify\authors
  \def\\{\protect\linebreak}%
  \textsc{\normalsize\textcolor{darkelectricblue}{\authors}}%
  \ifx\@empty\contribs
  \else
    ,\penalty-3 \space \@setcontribs
    \@closetoccontribs
  \fi
  \endtrivlist
  \endgroup
}
\def\@settitle{\begin{center}%
  \baselineskip14\p@\relax
    \bfseries
    \textsc{\Large\textcolor{charcoal}{\@title}}
  \end{center}%
}
\makeatother

\usepackage{enumitem}
\setlist[description]{%
  topsep=30pt,               % space before start / after end of list
  itemsep=5pt,               % space between items
  font={\bfseries\sffamily\color{NavyBlue}}, % if colour is needed
}

\usepackage{fancyhdr}
\usepackage{lastpage}

\newcommand*\Title{\textcolor{bluepigment}{Stable explicit schemes}}
\newcommand*\Authors{\textcolor{bluepigment}{S.~Gasparin, J.~Berger, D.~Dutykh \& N.~Mendes}}
\newcommand*{\plogo}{\textcolor{gray}{{\texttt{arXiv.org} / \textsc{hal}}}} % Generic publisher logo

\numberwithin{equation}{section}

\newcommand{\DF}{\textsc{Dufort}--\textsc{Frankel}}
\newcommand{\Eu}{\textsc{Euler}}
\newcommand{\CN}{\textsc{Crank}--\textsc{Nicolson}}
\newcommand{\mCN}{modified \textsc{Crank}--\textsc{Nicolson}}

\newcommand*\egal{\ = \ }
\newcommand*\plus{\ + \ }
\newcommand*\moins{\ - \ }

\newcommand{\f}{\mathrm{f}}
\renewcommand{\O}{\mathcal{O}}
\newcommand{\x}{\boldsymbol{x}}
\newcommand*{\Ox}{\Omega_{\, x}}
\newcommand*{\Ot}{\Omega_{\, t}}
\newcommand{\const}{\mathrm{const}}

\newcommand{\BivL}{\mathrm{Bi}_{\,v}^{\,\mathrm{L}}}
\newcommand{\BivR}{\mathrm{Bi}_{\,v}^{\,\mathrm{R}}}
\newcommand{\cm}{c_{\,m}}
\newcommand{\cms}{c_{\,m}^{\,\star}}
\newcommand{\dm}{d_{\,m}}
\newcommand{\dmref}{d_{\,m}^{\,0}}
\newcommand{\dms}{d_{\,m}^{\,\star}}

\newcommand{\glL}{g_{\,l}^{\,\mathrm{L}}}
\newcommand{\glR}{g_{\,l}^{\,R}}
\newcommand{\glsL}{g_{\,l, \,\mathrm{L}}^{\,\star}}     
\newcommand{\glsR}{g_{\,l,\, \mathrm{R}}^{\,\star}}

\newcommand{\hvL}{h_{\,v}^{\,\mathrm{L}}}
\newcommand{\hvR}{h_{\,v}^{\,\mathrm{R}}}
\newcommand{\kl}{k_{\,l}}
\newcommand{\kv}{k_{\,v}}
\newcommand{\Pc}{P_{\,c}}
\newcommand{\Ps}{P_{\,s}}
\newcommand{\Pv}{P_{\,v}}
\newcommand{\Pvi}{P_{\,v}^{\,i}}
\newcommand{\PvL}{P_{\,v}^{\,\mathrm{L}}}
\newcommand{\PvR}{P_{\,v}^{\,\mathrm{R}}}
\newcommand{\Rv}{R_{\,v}}
\newcommand{\tref}{t^{\,0}}
\newcommand{\uL}{u^{\,\mathrm{L}}}
\newcommand{\uR}{u^{\,\mathrm{R}}}
\newcommand{\rholv}{\rho_{\,l+v}}
\newcommand{\xs}{x^{\,\star}}
\newcommand{\ts}{t^{\,\star}}
\newcommand{\M}{\mathcal{M}}

\newcommand{\scal}{\boldsymbol{\cdot}}
\newcommand*\pd[2]{\frac{\partial #1}{\partial #2}}
\renewcommand{\div}{\grad\scal}
\newcommand{\grad}{\boldsymbol{\nabla}}
\newcommand{\eqdef}{\mathop{\stackrel{\,\mathrm{def}}{:=}\,}}
\newcommand{\norm}[1]{\lVert\, #1\, \rVert}
\renewcommand{\L}{\mathcal{L}}
\newcommand{\abs}[1]{\lvert\, #1\, \rvert}

\newcommand{\half}{{\textstyle{1\over2}}}
\newcommand{\sixth}{{\textstyle{1\over6}}}
\newcommand{\twelwth}{{\textstyle{1\over12}}}
\newcommand{\dix}[1]{ \cdot 10^{\,#1}}

\renewcommand{\Re}{\operatorname{Re}}
\renewcommand{\Im}{\operatorname{Im}}

\begin{document}

\title[\Title]{Stable explicit schemes for simulation of nonlinear moisture transfer in porous materials}

\author[S.~Gasparin]{Suelen Gasparin$^*$}
\address{\textbf{S.~Gasparin:} Thermal Systems Laboratory, Mechanical Engineering Graduate Program, Pontifical Catholic University of Paran\'a, Rua Imaculada Concei\c{c}\~{a}o, 1155, CEP: 80215-901, Curitiba -- Paran\'a, Brazil}
\email{suelengasparin@hotmail.com}
\urladdr{https://www.researchgate.net/profile/Suelen\_Gasparin/}
\thanks{$^*$ Corresponding author}

\author[J.~Berger]{Julien Berger}
\address{\textbf{J.~Berger:} Thermal Systems Laboratory, Mechanical Engineering Graduate Program, Pontifical Catholic University of Paran\'a, Rua Imaculada Concei\c{c}\~{a}o, 1155, CEP: 80215-901, Curitiba -- Paran\'a, Brazil}
\email{Julien.Berger@pucpr.edu.br}
\urladdr{https://www.researchgate.net/profile/Julien\_Berger3/}

\author[D.~Dutykh]{Denys Dutykh}
\address{\textbf{D.~Dutykh:} LAMA, UMR 5127 CNRS, Universit\'e Savoie Mont Blanc, Campus Scientifique, 73376 Le Bourget-du-Lac Cedex, France}
\email{Denys.Dutykh@univ-savoie.fr}
\urladdr{http://www.denys-dutykh.com/}

\author[N.~Mendes]{Nathan Mendes}
\address{\textbf{N.~Mendes:} Thermal Systems Laboratory, Mechanical Engineering Graduate Program, Pontifical Catholic University of Paran\'a, Rua Imaculada Concei\c{c}\~{a}o, 1155, CEP: 80215-901, Curitiba -- Paran\'a, Brazil}
\email{Nathan.Mendes@pucpr.edu.br}
\urladdr{https://www.researchgate.net/profile/Nathan\_Mendes/}


%%% ------------------------------------------------------------------------ %%%

\begin{titlepage}
\thispagestyle{empty} % Remove page numbering on this page
\noindent
{\Large Suelen \textsc{Gasparin}}\\
{\it\textcolor{gray}{Pontifical Catholic University of Paran\'a, Brazil}}
\\[0.02\textheight]
{\Large Julien \textsc{Berger}}\\
{\it\textcolor{gray}{Pontifical Catholic University of Paran\'a, Brazil}}
\\[0.02\textheight]
{\Large Denys \textsc{Dutykh}}\\
{\it\textcolor{gray}{CNRS, Universit\'e Savoie Mont Blanc, France}}
\\[0.02\textheight]
{\Large Nathan \textsc{Mendes}}\\
{\it\textcolor{gray}{Pontifical Catholic University of Paran\'a, Brazil}}
\\[0.08\textheight]

\colorbox{Lightblue}{
  \parbox[t]{1.0\textwidth}{
    \centering\huge\sc
    \vspace*{0.7cm}
    
    \textcolor{bluepigment}{Stable explicit schemes for simulation of nonlinear moisture transfer in porous materials}

    \vspace*{0.7cm}
  }
}

\vfill % Whitespace between the title block and the publisher

\raggedleft     % Right-align all text
{\large \plogo} % Publisher and logo
\end{titlepage}

%%% ------------------------------------------------------------------------ %%%

\newpage
\thispagestyle{empty} % Remove page numbering on this page
\par\vspace*{\fill}   % Whitespace until the bottom
\begin{flushright} % Right-align all text
{\textcolor{denimblue}{\textsc{Last modified:}} \today}
\end{flushright}

%%% ------------------------------------------------------------------------ %%%

\newpage
\maketitle
\thispagestyle{empty}

%%% ------------------------------------------------------------------------ %%%

\begin{abstract}

Implicit schemes have been extensively used in building physics to compute the solution of moisture diffusion problems in porous materials for improving stability conditions. Nevertheless, these schemes require important sub-iterations when treating nonlinear problems. To overcome this disadvantage, this paper explores the use of improved explicit schemes, such as \DF, \CN ~and hyperbolisation approaches. A first case study has been considered with the hypothesis of linear transfer. The \DF, \CN ~and hyperbolisation schemes were compared to the classical \Eu ~explicit scheme and to a reference solution. Results have shown that the hyperbolisation scheme has a stability condition higher than the standard \textsc{Courant-Friedrichs-Lewy} (CFL) condition. The error of this schemes depends on the parameter $\tau$ representing the hyperbolicity magnitude added into the equation. The \DF ~scheme has the advantages of being unconditionally stable and is preferable for nonlinear transfer, which is the three others cases studies. Results have shown the error is proportional to $\O(\Delta t)\,$. A \mCN ~scheme has been also studied in order to avoid sub-iterations to treat the nonlinearities at each time step. The main advantages of the \DF ~scheme are (i) to be twice faster than the \CN ~approach; (ii) to compute \emph{explicitly} the solution at each time step; (iii) to be unconditionally stable and (iv) easier to parallelise on high-performance computer systems. Although the approach is unconditionally stable, the choice of the time discretisation $\Delta t$ remains an important issue to accurately represent the physical phenomena.

\bigskip
\noindent \textbf{\keywordsname:} moisture diffusion; numerical methods; finite differences; explicit schemes; CFL condition; Dufort--Frankel scheme \\


\end{abstract}

%%% ------------------------------------------------------------------------ %%%

\newpage
\tableofcontents
\thispagestyle{empty}

%%% ------------------------------------------------------------------------ %%%

\newpage
\section{Introduction}

Excessive levels of moisture may lead to mould growth and may affect the indoor air quality, the thermal comfort of the occupants and HVAC energy consumption and demand. Moreover, it can deteriorate building fa\c{c}ades and decrease envelope durability \cite{Harris2001, Berger2015a}. Assessment of relative humidity in constructions is also important for management and performance of HVAC systems.

Models for moisture transfer in porous building materials have been implemented as building simulation tools since the nineties in software such as \textsc{Delphin} \cite{BauklimatikDresden2011}, \textsc{MATCH} \cite{Rode2003}, \textsc{MOIST} \cite{Burch1993}, \textsc{WUFI} \cite{IBP2005} and \textsc{UMIDUS} \cite{Mendes1997, Mendes1999} among others. Moisture models have also been implemented in whole-building simulation tools and tested in the frame of the International Energy Agency Annex~$41$, which reported on most of detailed models and their successful applications for accurate assessment of hygrothermal transfer in buildings \cite{Woloszyn2008}.

Moisture transfer is represented by a diffusion equation, formulated as: 
\begin{align*}
  \pd{u}{t} \egal \div \left( \, \nu\, \grad u \, \right) \,,
\end{align*}
associated to boundary and initial conditions, where $\nu$ is the diffusion of the material and where $u\,(\x,t)$ is the moisture potential being diffused in the spatial domain $\Ox$ during the time interval $\Ot$.We denote $\Delta x$ and $\Delta t$ the spatial and time discretisation within those the domains. Due to the nonlinearities of the material properties and due to the non-periodicity of the boundary conditions, the models use numerical techniques to compute the moisture content from the partial differential governing equation.

Due to its property of unconditional stability, the \Eu ~implicit scheme has been used in many works reported in the literature \cite{Mendes2005, IBP2005, BauklimatikDresden2011, Steeman2009, Rouchier2013, Janssen2014, Janssen2007}. However, it has the order of accuracy $\O\,(\Delta t\ +\ \Delta x^{\,2})\,$, while the \textsc{Crank--Nicolson} scheme can be used to increase the accuracy to $\O\,(\Delta t^{\,2} \ +\ \Delta x^{\,2})\,$. For the same time and space discretisation, numerical results obtained with this scheme are more accurate than those obtained from \Eu ~implicit scheme. The \CN ~scheme has been implemented for instance in \cite{VanGenuchten1982}. Nevertheless, at every time step, one has to use a tridiagonal solver to invert the linear system of equations to determine the solution value at the following time layer. For instance in \cite{Mendes2005}, a multi-tridiagonal matrix algorithm has been developed to compute the solution of coupled equations of nonlinear heat and moisture transfer, using an \Eu ~implicit scheme. Furthermore, when dealing with nonlinearities of the material properties for instance, one has to perform sub-iterations to linearise the system, increasing thus the total number of iterations. In \cite{Janssen2014}, thousands of iterations are required to converge to the solution of a mass diffusion problem.

On the other hand, an explicit scheme enables a direct computation of the the solution at the following time layer. Some examples of works based on explicit schemes can be found in the literature as \cite{Tariku2010, Kalagasidis2007}. Nevertheless, this scheme is conditionally stable under the \textsc{Courant--Friedrichs--Lewy} (CFL) condition:
\begin{align*}
  \Delta t\ \leqslant\ \frac{1}{2\,\nu}\;\Delta x^{\,2} \,.
\end{align*}
The CFL condition is restrictive for fine discretisations, explaining why few works have used this approach in building physics.

This paper is devoted to explore the use of improved explicit schemes to overcome the instability limitation of the standard explicit scheme. The proposed schemes are evaluated to solve nonlinear transfer of moisture in porous material. The advantages and drawbacks are discussed for two numerical applications. Next Section aims at describing the physical phenomena of moisture transfer in porous material. In Section~\ref{sec:numerical_schemes}, basics of the \DF ~and the hyperbolisation explicit schemes, are detailed. Then, linear and nonlinear moisture transfer cases are considered to verify the features of the proposed approaches.

%%% ------------------------------------------------------------------------ %%%

\section{Moisture transfer in porous materials}

The physical problem involves one-dimension moisture diffusion through a porous material defined by the spatial domain $\Ox \egal [\, 0, \, L \,] $. The moisture transfer occurs according to the liquid and vapour diffusion. The physical problem can be formulated as \cite{Janssen2014}:
\begin{align}\label{eq:moisture_equation_1D}
  & \pd{\rholv}{t} \ = \ \pd{}{x} \left( \, \kl \, \pd{\Pc}{x} \plus \kv \, \pd{\Pv}{x} \, \right) \,,
\end{align}
where $\rholv$ is the volumetric moisture content of the material and $\kv$ and $\kl$, the vapour and liquid permeabilities.

Eq.~\eqref{eq:moisture_equation_1D} can be written using the vapour pressure $\Pv$ as the driving potential. For this, we consider the physical relation, known as the \textsc{Kelvin} equation, between $\Pv$ and $\Pc$:
\begin{align*}
  \Pc & \egal \Rv \cdot T \cdot \ln\left(\frac{\Pv}{\Ps(T)}\right) \\
  \pd{\Pc}{\Pv} & \egal \frac{R_{\,v} \, T}{\Pv} \,.
\end{align*}
Thus we have:
\begin{align*}
  \pd{\Pc}{x} \egal \pd{\Pc}{\Pv} \cdot \pd{\Pv}{x} \plus \pd{\Pc}{T} \cdot \pd{T}{x} \,.
\end{align*}
As we consider the mass transfer under isothermal conditions, the second term vanishes and we obtain: 
\begin{align*}
  \pd{\Pc}{x} \egal \frac{\Rv \, T}{\Pv} \cdot \pd{\Pv}{x} \,.
\end{align*}
In addition, we have:
\begin{align*}
  \pd{\rholv}{t} \egal \pd{\rholv}{\phi} \cdot \pd{\phi}{\Pv} \cdot \pd{\Pv}{t} \plus \pd{\rholv}{T} \cdot \pd{T}{t} \simeq \pd{\rholv}{\phi} \cdot \pd{\phi}{\Pv} \cdot \pd{\Pv}{t} \,.
\end{align*}
Considering the relation $\rholv \egal \f(\phi) \egal \f(\Pv,T)\,$, obtained from material properties and from the relation between the vapour pressure $\Pv$ and the relative humidity $\phi\,$, we get:
\begin{align*}
  \pd{\rholv}{t} \egal \f^{\,\prime}(\Pv) \; \frac{1}{\Ps} \; \pd{\Pv}{t} \,.
\end{align*}
Eq.~\eqref{eq:moisture_equation_1D} can be therefore rewritten as:
\begin{align}\label{eq:moisture_equation_1D_v2}
  & \f^{\,\prime}(\Pv) \; \frac{1}{\Ps} \; \pd{\Pv}{t} \egal \pd{}{x} \biggl[ \, \Bigl( \, \kl \, \frac{\Rv \, T}{\Pv} \plus \kv \, \Bigr) \, \pd{\Pv}{x} \, \biggr] \,.
\end{align}
The material properties $\f^{\,\prime}(\Pv)$, $\kl$ and $\kv$ depend on the vapour pressure $\Pv$. At the material bounding surfaces, \textsc{Robin}-type boundary conditions are considered:
\begin{align}\label{eq:bc}
  \left( \, \kl \, \frac{\Rv \, T}{\Pv} \plus \kv \, \right) \, \pd{\Pv}{x} &\egal 
\hvL \cdot \left( \, \Pv \moins \PvL \, \right) \moins \glL \, , && x \egal 0 \,, \\
  - \left( \, \kl \, \frac{\Rv \, T}{\Pv} \plus \kv \, \right) \, \pd{\Pv}{x} &\egal \hvR \cdot \left( \, \Pv \moins \PvR \, \right) \moins \glR \, ,&& x \egal L \,,
\end{align}
where $\PvL$ and $\PvR$ are the vapour pressure of the ambient air, $\glL$ and $\glR$ are the liquid flow (driving rain) at the two bounding surfaces. We consider a uniform vapour pressure distribution as initial condition:
\begin{align}\label{eq:ic}
  \Pv &\egal \Pvi \,, && t\egal0 \,.
\end{align}
While performing a mathematical and numerical analysis of a given practical problem, it is of capital importance to obtain a unitless formulation of governing equations, due to a number of good reasons. First of all, it enables to determine important scaling parameters (\textsc{Biot} numbers for instance). Henceforth, solving one dimensionless problem is equivalent to solve a whole class of dimensional problems sharing the same scaling parameters. Then, dimensionless equations allow to estimate the relative magnitude of various terms, and thus, eventually to simplify the problem using asymptotic methods \cite{Nayfeh2000}. Finally, the floating point arithmetics is designed such as the rounding errors are minimal if you manipulate the numbers of the same magnitude \cite{Kahan1979}. Moreover, the floating point numbers have the highest density in the interval $(0,\,1)$ and their density decays exponentially when we move further away from zero. So, it is always better to manipulate numerically the quantities at the order of $\O(1)$ to avoid severe round-off errors and to likely improve the conditioning of the problem in hands. Therefore, we denote $\dm \egal \kl \cdot \dfrac{\Rv \, T}{\Pv} \plus \kv $ as a global moisture transport coefficient, $\cm \egal \f^{\,\prime}(\Pv) \; \dfrac{1}{\Ps}$ as the moisture storage coefficient and define the following dimensionless quantities:
\begin{align*}
& u \egal \frac{\Pv}{\Pvi} \,,
&& \uR \egal \frac{\PvR}{\Pvi} \,,
&& \uL \egal \frac{\PvL}{\Pvi} \,,
&& \xs \egal \frac{x}{L} \,, \\[3pt]
& \ts \egal \frac{t}{\tref} \,,
&& \cms \egal \frac{\cm \cdot L^2}{\dmref \cdot \tref} \,,
&& \dms \egal \frac{\dm}{\dmref} \,,
&& \BivL \egal \frac{\hvL \cdot L}{\dmref}  \,,\\[3pt]
& \BivR \egal \frac{\hvR \cdot L}{\dmref} \,,
&& \glsL \egal \frac{\glL \cdot L}{\dmref \cdot \Pvi} \,,
&& \glsR \egal \frac{\glR \cdot L}{\dmref \cdot \Pvi}\,.
\end{align*}
In this way, the dimensionless governing equations are then written as:
\begin{subequations}\label{eq:moisture_dimensionlesspb_1D}
\begin{align}
  \cms \pd{u}{\ts} &\egal \pd{}{\xs} \left( \, \dms \pd{u}{\xs} \, \right) \,,
& \ts & \ > \ 0\,, \;&  \xs & \ \in \ \big[ \, 0, \, 1 \, \big] \,, \\[3pt]
  \dms \, \pd{u}{\xs} &\egal \BivL \cdot \left( \, u \moins \uL \, \right) \moins \glsL \,,
& \ts & \ > \ 0\,, \,&  \xs & \egal 0 \,, \\[3pt]
 -\dms \, \pd{u}{\xs} &\egal \BivR \cdot \left( \, u \moins \uR \, \right) \moins \glsR \,,
& \ts & \ > \ 0\,, \,&   \xs & \egal 1 \,, \\[3pt]
 u &\egal 1 \,,
& \ts & \egal 0\,, \,&  \xs & \ \in \ \big[ \, 0, \, 1 \, \big] \,.
\end{align}
\end{subequations}

%%% ------------------------------------------------------------------------ %%%

\section{Numerical schemes}
\label{sec:numerical_schemes}

In order to describe numerical schemes, let us consider a uniform discretisation of the interval $\Ox \ \rightsquigarrow\ \Omega_{\,h}\,$:
\begin{equation*}
  \Omega_{\,h}\ =\ \bigcup_{j\,=\,0}^{N-1} [\,x_{\,j},\;x_{\,j+1}\,]\,, \qquad
  x_{j+1}\ -\ x_{\,j}\ \equiv\ \Delta x\,, \quad \forall j\ \in\ \bigl\{0,\,1,\,\ldots,\,N-1\bigr\}\,.
\end{equation*}
The time layers are uniformly spaced as well $t^{\,n}\ =\ n\,\Delta t\,$, $\Delta t\ =\ \const\ >\ 0\,$, $n\ =\ 0,\,1,\,2,\,\ldots, \, N_{\,t}$ The values of function $u(x,\,t)$ in discrete nodes will be denoted by $u_{\,j}^{\,n}\ \eqdef\ u\,(x_{\,j},\,t^{\,n}\,)\,$.

For the sake of simplicity and without loosing the generality, the numerical schemes are explained considering $\dms$ and $\cms$ as constant, noting $\nu = \dfrac{\dms}{\cms}$ and the linear diffusion equation:
\begin{align}\label{eq:heat1d}
  \pd{u}{t} \egal \div \left( \, \nu\, \grad u \, \right)\,.
\end{align}

\begin{figure}[h!]
\begin{center}
\subfigure[][\label{fig:stencil_explicit} Euler Explicit]{\includegraphics[width=0.48\textwidth]{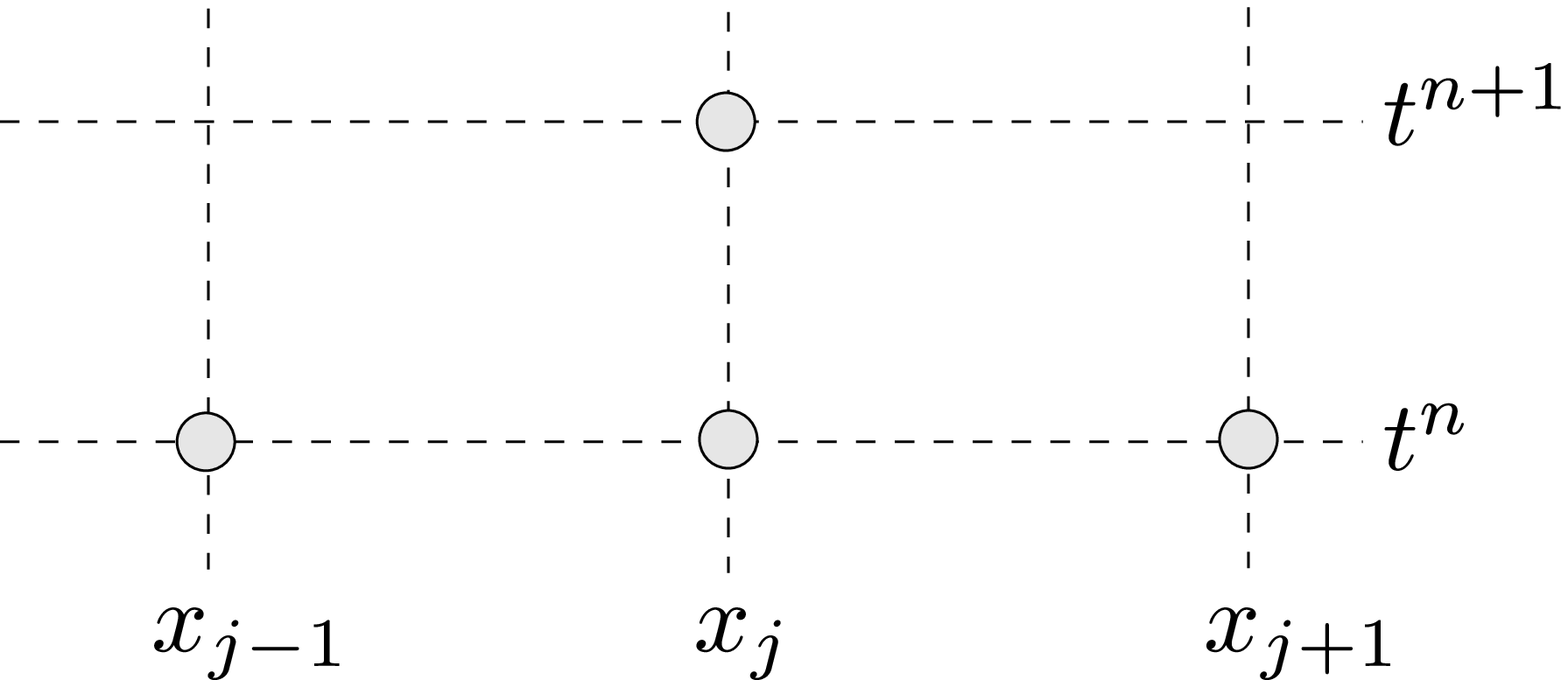}}
\subfigure[][\label{fig:stencil_CN} Crank--Nicolson]{\includegraphics[width=0.48\textwidth]{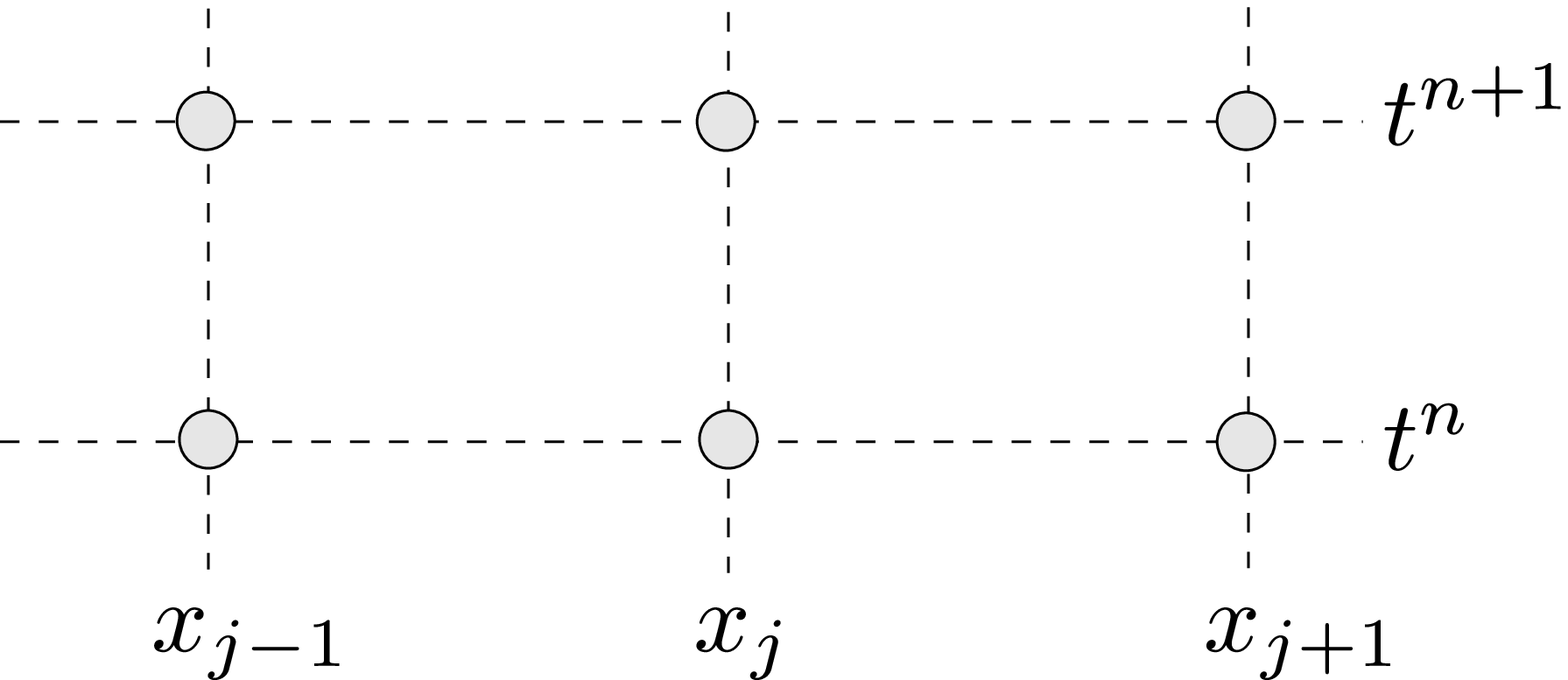}}
\subfigure[][\label{fig:stencil_Dufort} Dufort--Frankel]{\includegraphics[width=0.48\textwidth]{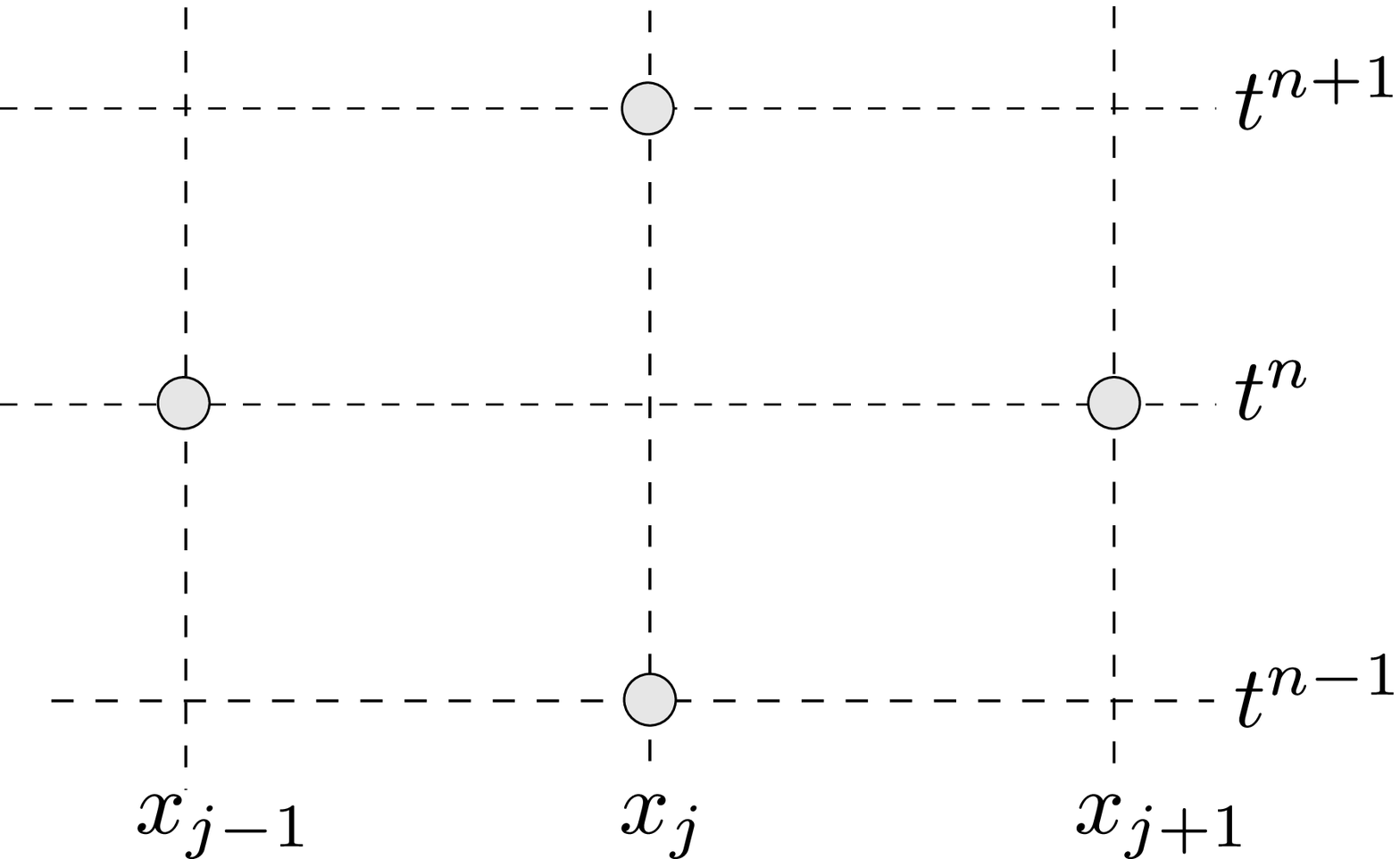}}
\subfigure[][\label{fig:stencil_leap} Hyperbolisation]{\includegraphics[width=0.48\textwidth]{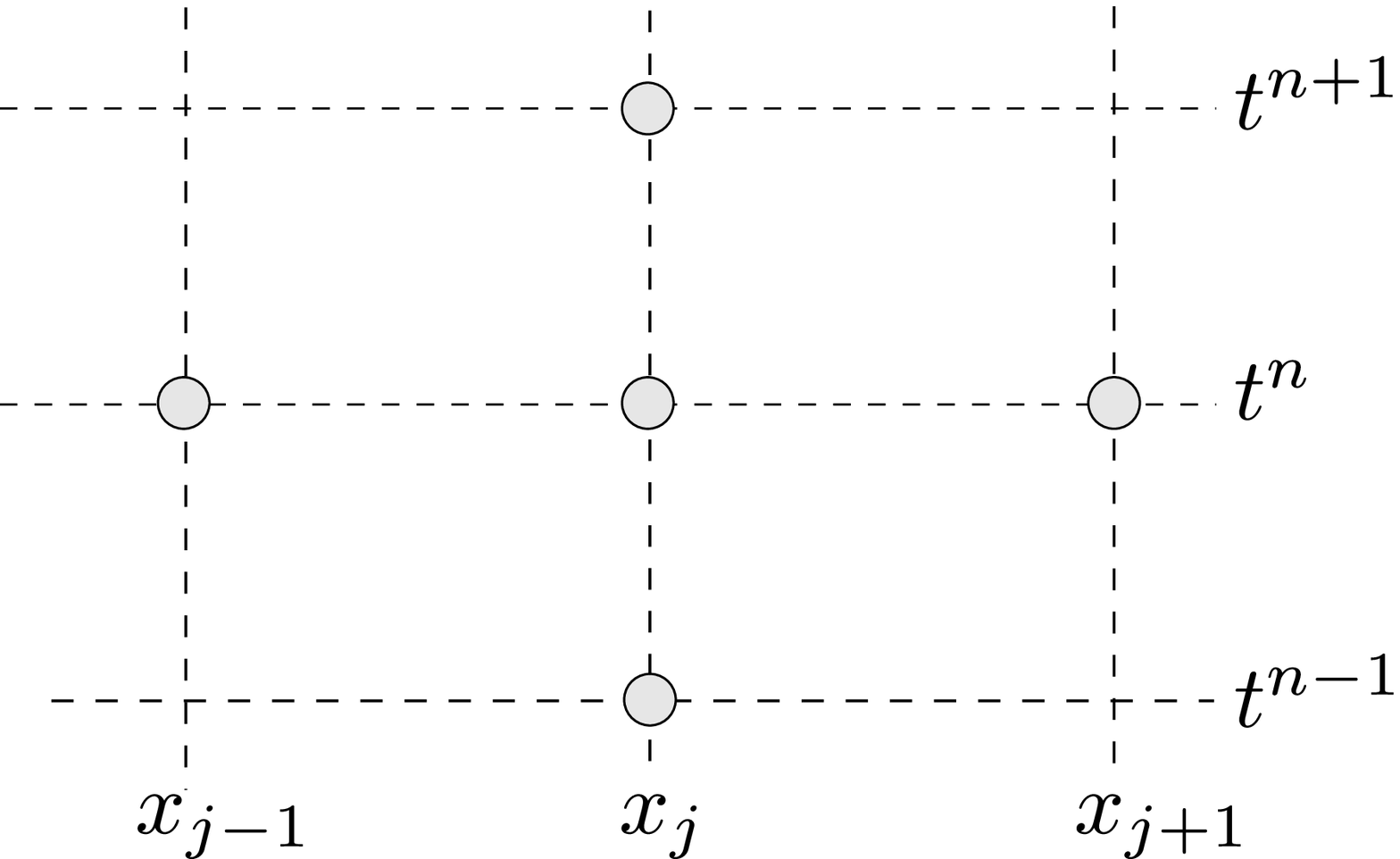}}
\caption{\small\em Stencils of the numerical schemes.}
\label{fig:stencils}
\end{center}
\end{figure}

%%% ------------------------------------------------------------------------ %%%

\subsection{The \CN ~scheme}
\label{sec:crank}

A very useful method was proposed by \textsc{Crank} \& \textsc{Nicolson} (CN) and it can be successfully applied to the diffusion equation \eqref{eq:heat1d} as well:
\begin{multline}\label{eq:cn}
  \frac{u_{\,j}^{\,n+1}\ -\ u_{\,j}^{\,n}}{\Delta t}\ =\ \nu\;\frac{u_{\,j-1}^{\,n}\ -\ 2\,u_{\,j}^{\,n}\ +\ u_{\,j+1}^{\,n}}{2\,\Delta x^{\,2}}\ +\ \nu\;\frac{u_{\,j-1}^{\,n+1}\ -\ 2\,u_{\,j}^{\,n+1}\ +\ u_{\,j+1}^{\,n+1}}{2\,\Delta x^{\,2}}\,, \\ j\ =\ 1,\,\ldots,\,N-1\,, \qquad n\ \geq\ 0\,.
\end{multline}
This scheme is $\O(\Delta t^2\ +\ \Delta x^{\,2})$ accurate and unconditionally stable. That is why numerical results obtained with the CN scheme will be more accurate than implicit scheme predictions. The stencil of this scheme is depicted in Figure~\ref{fig:stencil_CN}. The CN scheme has all advantages and disadvantages (except for the order of accuracy in time) of the implicit scheme. At every time step one has to use a tridiagonal solver to invert the linear system of equations to determine solution value at the following time layer $t\ =\ t^{\,n+1}\,$.

%%% ------------------------------------------------------------------------ %%%

\subsection{The \Eu ~explicit scheme}

The standard explicit scheme can be written as:
\begin{align}\label{eq:exp}
& \frac{u_{\,j}^{\,n+1}\ -\ u_{\,j}^{\,n}}{\Delta t}\ =\ \nu\;\frac{u_{\,j-1}^{\,n}\ -\ 2\,u_{\,j}^{\,n}\ +\ u_{\,j+1}^{\,n}}{\Delta x^{\,2}}\,, \qquad j\ =\ 1,\,\ldots,\,N-1\,, \qquad n\ \geqslant\ 0\,.
\end{align}
The stencil of this scheme is depicted in Figure~\ref{fig:stencil_explicit}. This discretisation is completed using the two boundary conditions: 
\begin{align*}
& u_{\,0}^{\,n+1}\ =\ \psi_{\,\mathrm{L}}\,(t^{\,n+1},\,u_{\,1}^{\,n+1},\,\ldots\,) \,,\\
& u_{\,N}^{\,n+1}\ =\ \psi_{\,\mathrm{R}}\,(t^{\,n+1},\,u_{\,N-1}^{\,n+1},\,\ldots\,) \,,
\end{align*}
where functions $\psi_{\,\mathrm{l},\,\mathrm{r}}\,(\,\bullet\,)$ may depend on adjacent values of the solution whose number depends on the approximation order of the scheme (here we use the second order in space). For instance, for the left boundary conditions, we have 
\begin{align*}
  \dms\;\frac{-3\,u_{\,0}^{\,n+1}\ +\ 4\,u_{\,1}^{\,n+1}\ -\ u_{\,2}^{\,n+1}}{2\,\Delta x} \egal \BivL \cdot \left(\, u_{\,0}^{\,n+1} \moins \uL \, \right) \moins \glsL \,.
\end{align*}

By solving Eq.~\eqref{eq:exp} with respect to $u_{\,j}^{\,n+1}$, we obtain a discrete dynamical system
\begin{align*}
  u_{\,j}^{\,n+1}\ =\ u_{\,j}^{\,n}\ +\ \nu\;\frac{\Delta t}{\Delta x^{\,2}}\;\bigl( \, u_{\,j-1}^{\,n}\ -\ 2\,u_{\,j}^{\,n}\ +\ u_{\,j+1}^{\,n} \, \bigr) \,,
\end{align*}
whose starting value is directly obtained from the initial condition:
\begin{align*}
  u_{\,j}^{\,0}\ =\ 1 \,.
\end{align*}
It is well-known that scheme \eqref{eq:exp} approximates the continuous operator to order $\O(\Delta t\ +\ \Delta x^{\,2})\,$. The explicit scheme is conditionally stable under the following CFL-type condition:
\begin{align}\label{eq:cfl}
  \Delta t\ \leqslant\ \frac{1}{2\,\nu}\;\Delta x^{\,2} \,.
\end{align}
Unfortunately, this condition is too restrictive for sufficiently fine discretisations.

%%% ------------------------------------------------------------------------ %%%

\subsection{Improved explicit scheme: \DF ~method}

Using the so-called \textsc{Dufort}--\textsc{Frankel} method, the numerical scheme is expressed as:
\begin{align}\label{eq:dufort}
  & \frac{u_{\,j}^{\,n+1}\ -\ u_{\,j}^{\,n-1}}{2\,\Delta t}\ =\ \nu\;\frac{u_{\,j-1}^{\,n}\ -\ \bigl(u_{\,j}^{\,n-1}\ +\ u_{\,j}^{\,n+1}\bigr)\ +\ u_{\,j+1}^{\,n}}{\Delta x^{\,2}}\,, \qquad j\ =\ 1,\,\ldots,\,N-1\,, \qquad n\ \geqslant\ 0\,,
\end{align}
where the term $2\,u_{\,j}^{\,n}\ $ is replaced by $u_{\,j}^{\,n-1}\ +\ u_{\,j}^{\,n+1}$. The scheme \eqref{eq:dufort} has the stencil depicted in Figure~\ref{fig:stencil_Dufort}. At first glance, the scheme \eqref{eq:dufort} looks like an implicit scheme, however, it is not truly the case. Eq.~\eqref{eq:dufort} can be easily solved for $u_{\,j}^{\,n+1}$ to give the following discrete dynamical system:
\begin{align*}
  u_{\,j}^{\,n+1}\ =\ \frac{1\ -\ \lambda}{1\ +\ \lambda}\;u_{\,j}^{\,n-1}\ +\ \frac{\lambda}{1\ +\ \lambda}\;\bigl(u_{\,j+1}^{\,n}\ +\ u_{\,j-1}^{\,n}\bigr) \,,
\end{align*}
where:
\begin{align*}
  \lambda\ \eqdef\ 2\,\nu\;\frac{\Delta t}{\Delta x^{\,2}} \,.
\end{align*}
The standard \textsc{von Neumann} stability analysis (detailed in the Appendix) shows that the \textsc{Dufort}--\textsc{Frankel} scheme is \emph{unconditionally stable}.

The consistency error analysis of the scheme \eqref{eq:dufort} shows the following result:
\begin{multline}\label{eq:expansion_dufort}
  \L_{\,j}^{\,n} \egal \nu\;\frac{\Delta t^{\,2}}{\Delta x^{\,2}} \;\pd{^2u}{t^2} \plus \pd{u}{t} \ -\ \nu\,\pd{^2u}{x^2} \plus   \sixth\,\Delta t^{\,2}\,\pd{^3u}{t^3} \\
 \moins \twelwth\,\nu\,\Delta x^{\,2}\,\pd{^4u}{x^4} \ -\ \twelwth\,\nu\,\Delta t^{\,2}\,\Delta x\,\pd{^3}{x^3}\pd{^2u}{t^2}\ +\ \O\Bigl(\frac{\Delta t^{\,4}}{\Delta x^{\,2}}\Bigr) \,,
\end{multline}
where
\begin{align*}
\L_{\,j}^{\,n}\ \eqdef\ \frac{u_{\,j}^{\,n+1}\ -\ u_{\,j}^{\,n-1}}{2\,\Delta t}\ -\ \nu\;\frac{u_{\,j-1}^{\,n}\ -\ \bigl(u_{\,j}^{\,n-1}\ +\ u_{\,j}^{\,n+1}\bigr)\ +\ u_{\,j+1}^{\,n}}{\Delta x^{\,2}} \,.
\end{align*}
So, from the asymptotic expansion for $\L_{\,j}^{\,n}$ we obtain that the \textsc{Dufort}--\textsc{Frankel} scheme is second order accurate in time and:
\begin{itemize}
  \item First order accurate in space if $\Delta t\ \propto\ \Delta x^{\,3/2}$
  \item Second order accurate in space if $\Delta t\ \propto\ \Delta x^{\,2}$
\end{itemize}
It is important to note that the boundary condition have to be discretised to the second order of accuracy $\O(\Delta x^{\,2})$ to maintain the method features uniformly.

%%% ------------------------------------------------------------------------ %%%

\subsection{Hyperbolisation scheme}

From Eq.~\eqref{eq:expansion_dufort}, it can be noted that the \textsc{Dufort}--\textsc{Frankel} scheme is unconditionally consistent with the so-called \emph{hyperbolic heat conduction equation}. Thus, the scheme is a hidden way to add a small amount of 'hyperbolicity' into the model \eqref{eq:heat1d}. In this Section we shall invert the order of operations: first, we perturb Eq.~\eqref{eq:heat1d} in an ad-hoc way and only after we discretise it with a suitable method. We consider the $1-$dimension Eq.~\eqref{eq:heat1d} perturbed by adding a low magnitude term containing the second derivative in time:
\begin{align}\label{eq:hyp}
  &  \tau\,\pd{^2u}{t^2} \plus \pd{u}{t} \egal \nu\,\pd{^2u}{x^2}  \,.
\end{align}

This is the \emph{hyperbolic diffusion equation} already familiar to us since it appeared in the consistency analysis of the \textsc{Dufort}--\textsc{Frankel} scheme. Here we perform a singular perturbation by assuming that
\begin{align*}
  \norm{\tau\,\pd{^2u}{t^2}}\ \ll\ \norm{\pd{u}{t}}\,.
\end{align*}
The last condition physically means that the new term has only limited influence on the solution of Eq.~\eqref{eq:hyp}. Here $\tau$ is a small ad-hoc parameter whose value is in general related to physical and discretisation parameters $\tau\ =\ \tau\,(\nu,\,\Delta x,\,\Delta t)\,$.

One can notice Eq.~\eqref{eq:hyp} requires two initial conditions to obtain a well-posed initial value problem. However, the parabolic Eq.~\eqref{eq:heat1d} is only first order in time and it only requires the knowledge of the initial temperature field distribution. When we solve the hyperbolic Eq.~\eqref{eq:hyp}, the missing initial condition is simply chosen to be
\begin{align*}
  &  \pd{u}{t}  \egal 0 \,, && t \egal 0\,.
\end{align*}

%%% ------------------------------------------------------------------------ %%%

\subsubsection{Dispersion relation analysis}

The classical dispersion relation analysis looks at plane wave solutions:
\begin{equation}\label{eq:ans}
  u(x,\,t)\ =\ u_{\,0}\,\mathrm{e}^{\,\mathrm{i}\,(\kappa\,x\ -\ \omega\,t)}\,.
\end{equation}
By substituting this solution ansatz into Eq.~\eqref{eq:heat1d} we obtain the following relation between wave frequency $\omega$ and wavenumber $k\,$:
\begin{equation}\label{eq:disp}
  \omega(\kappa)\ =\ -\,\mathrm{i}\,\nu\,\kappa^{\,2}\,.
\end{equation}
The last relation is called the \emph{dispersion relation} even if the diffusion Eq.~\eqref{eq:heat1d} is not dispersive but dissipative. The real part of $\omega$ contains information about wave propagation properties (dispersive if $\ \frac{\Re\omega(\kappa)}{\kappa}\ \neq\ \const$ and non-dispersive otherwise) while the imaginary part describes how different modes $\kappa$ dissipate (if $\ \Im\omega(\kappa)\ <\ 0$) or grow (if $\ \Im\omega(\kappa)\ >\ 0\,$). The dispersion relation \eqref{eq:disp} gives the damping rate of different modes.

The same plane wave ansatz \eqref{eq:ans} can be substituted into the hyperbolic heat Eq. \eqref{eq:hyp} as well to give the following \emph{implicit} relation for the wave frequency $\omega\,$:
\begin{equation*}
  -\,\tau\,\omega^{\,2}\ -\ \mathrm{i}\,\omega\ +\ \nu\,\kappa^{\,2}\ =\ 0\,.
\end{equation*}
By solving this quadratic equation with complex coefficients for $\omega\,$, we obtain two branches:
\begin{equation*}
  \omega_{\,\pm}\,(\kappa)\ =\ \frac{-\,\mathrm{i}\ \pm\ \sqrt{4\,\nu\,\kappa^{\,2}\,\tau\ -\ 1}}{2\,\tau}\,.
\end{equation*}
This dispersion relation will be analysed asymptotically with $\tau\ \ll\ 1$ being the small parameter. The branch $\omega_{\,-}\,(\kappa)$ is not of much interest to us since it is constantly damped, \emph{i.e.}
\begin{equation*}
  \omega_{\,-}\,(\kappa)\ =\ -\;\frac{\mathrm{i}}{\tau}\ +\ \O(1)\,.
\end{equation*}
It is much more instructive to look at the positive branch $\omega_{\,+}\,(\kappa)\,$:
\begin{equation*}
  \omega_{\,+}\,(\kappa)\ =\ -\,\mathrm{i}\,\nu\,\kappa^{\,2}\,\bigl[\,1\ +\ \nu\,\kappa^{\,2}\,\tau\ +\ 2\,\nu^{\,2}\,\kappa^{\,4}\,\tau^{\,2}\ +\ \O(\tau^{\,3})\,\bigr]\,.
\end{equation*}
The last asymptotic expansion shows that for small values of parameter $\tau\,$, we obtain a valid asymptotic approximation of the dispersion relation \eqref{eq:disp} for the diffusion equation \eqref{eq:heat1d}.

%%% ------------------------------------------------------------------------ %%%

\subsubsection{Error estimate}

It is legitimate to ask the question how far the solutions $u_{\,h}\,(x,\,t)$ of the hyperbolic equation \eqref{eq:hyp} are from the solutions $u_{\,p}\,(x,\,t)$ of the parabolic diffusion equation \eqref{eq:heat1d} (for the same initial condition). This question for the initial value problem was studied in \cite{Myshetskaya2015} and we shall provide here only the obtained error estimate. Let us introduce the difference between two solutions:
\begin{equation*}
  \delta u\,(x,\,t)\ \eqdef\ u_{\,h}\,(x,\,t)\ -\ u_{\,p}\,(x,\,t)\,.
\end{equation*}
Then, the following estimate holds
\begin{equation*}
  \abs{\delta u\,(x,\,t)}\ \leq\ \tau\,\M\,\Bigl(1\ +\ \frac{2}{\sqrt{\pi}}\Bigr)\,\biggl(8\,\sqrt{2}\,\tau\ +\ \frac{\sqrt[4]{2\,\pi^2}}{2}\;T\biggr)\,,
\end{equation*}
where $T\ >\ 0$ is the time horizon and
\begin{equation*}
  \M\ \eqdef\ \sup_{\Omega_{\,\xi,\,\zeta}}\;\Bigl\vert\,\pd{^{\,2}u_{\,p}}{\,t^{\,2}}(\xi,\,\zeta)\,\Bigr\vert\,,
\end{equation*}
and the domain $\Omega_{\,\xi,\,\zeta}$ is defined as
\begin{equation*}
  \Omega_{\,\xi,\,\zeta}\ \eqdef\ \Bigl\{\,(\xi,\,\zeta)\ :\ 0\ \leq\ \zeta\ \leq\ t\,, \quad
  x\ -\ \frac{t\ -\ \zeta}{\sqrt{\tau}}\ \leq\ \xi\ \leq\ x\ +\ \frac{t\ -\ \zeta}{\sqrt{\tau}}\,\Bigr\}\,.
\end{equation*}

%%% ------------------------------------------------------------------------ %%%

\subsubsection{Discretisation}

Eq.~\eqref{eq:hyp} is discretised on the stencil depicted in Figure~\ref{fig:stencil_leap}:
\begin{multline}\label{eq:hyps}
  \L_{\,j}^{\,n}\ \eqdef\ \tau\;\frac{u_{\,j}^{\,n+1}\ -\ 2\,u_{\,j}^{\,n}\ +\ u_{\,j}^{\,n-1}}{\Delta t^{\,2}}\ +\ \frac{u_{\,j}^{\,n+1}\ -\ u_{\,j}^{\,n-1}}{2\,\Delta t}\ -\ \nu\;\frac{u_{\,j+1}^{\,n}\ -\ 2\,u_{\,j}^{\,n}\ +\ u_{\,j-1}^{\,n}}{\Delta x^{\,2}}\ =\ 0\,,\\ \qquad j\ =\ 1,\,\ldots,\,N-1\,, \qquad n\ \geqslant\ 0\,,
\end{multline}
Using the standard \textsc{Taylor} expansions, it can be proven that the scheme is consistent with hyperbolic heat Eq.~\eqref{eq:hyp} to the second order in space and in time $\O(\Delta t^{\,2}\ +\ \Delta x^{\,2})\,$.

The stability of the scheme \eqref{eq:hyps} was studied in \cite{Chetverushkin2012} and the following stability condition was obtained:
\begin{align}\label{eq:cond_hyp}
  \frac{\Delta t}{\Delta x}\ \leqslant\ \sqrt{\frac{\tau}{\nu}}\,.
\end{align}
The choice of parameter $\tau$ is therefore an important issue and will be discussed in next Sections.

%%% ------------------------------------------------------------------------ %%%

\subsection{Validation of the numerical solution}

One possible comparison of the numerical schemes can be done by computing the $\mathcal{L}_{\,\infty}$ error between the solution $u_{\, \mathrm{num}}$ and a reference solution $u_{\, \mathrm{ref}}\,$:
\begin{subequations}
\begin{align}
  & \varepsilon \ \eqdef\ \bigm| \bigl| \, u_{\,\mathrm{ref}} \moins u_{\,\mathrm{num}}\,\bigr| \bigr|_{\,\infty} 
\end{align}
\end{subequations}
The computation of the reference solution is detailed in further Sections.

%%% ------------------------------------------------------------------------ %%%

\section{Numerical application: linear transfer}
\label{sec:case_linear}

\begin{figure}
  \centering
  \includegraphics[width=0.56\textwidth]{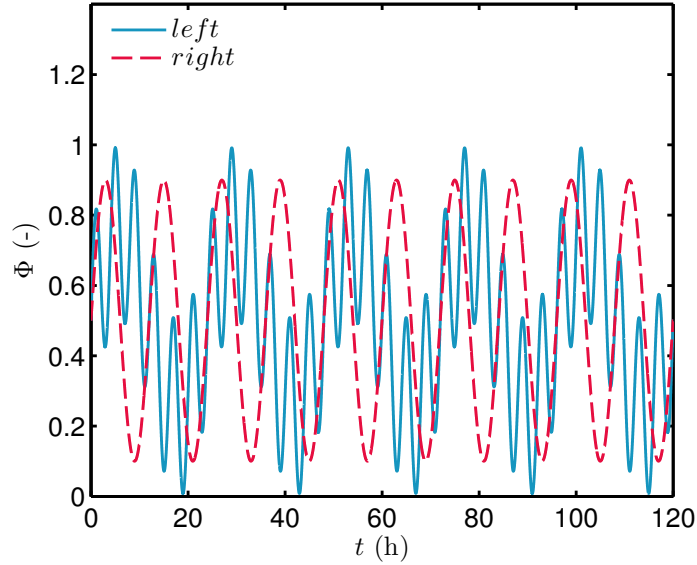}
  \caption{\small\em Boundary conditions.}
  \label{fig_AN1:BC}
\end{figure}

A first case of linear moisture transfer is considered. From a physical point of view, the numerical values correspond to a material length $L \egal 0.1$ $\mathsf{m}$. The moisture properties are $\dm \egal 1.97 \cdot 10^{-10}$ $\mathsf{s}$ and $\cm \egal 7.09 \cdot 10^{-3}$ $\mathsf{kg/m^3/Pa}$. The initial vapour pressure in the material is considered uniform $\Pvi \egal 1.16 \cdot 10^3$ $\mathsf{Pa \,}$, corresponding to a relative humidity of $50$\%. The reference time is $\tref \egal 1$  $\mathsf{h}$, thus the total time of simulation corresponds to 120 hours, or five days. The boundary conditions, represented by the relative humidity $\phi$ are given in Figure~\ref{fig_AN1:BC}. The sinusoidal variations oscillates between dry and moist state during the 120 hours. The convective vapour coefficients are set to $2 \cdot 10^{-7}$  $\mathsf{s/m}$ and $3 \cdot 10^{-8}$  $\mathsf{s/m}$ for the left and right boundary conditions, respectively. As the readers may be interested in testing the numerical schemes proposed,  in the present paper dimensionless values are provided in Appendix~\ref{annexe:dimensionless}.

\begin{figure}
  \centering
  \subfigure[a][\label{fig_AN1:x0time}]{\includegraphics[width=0.48\textwidth]{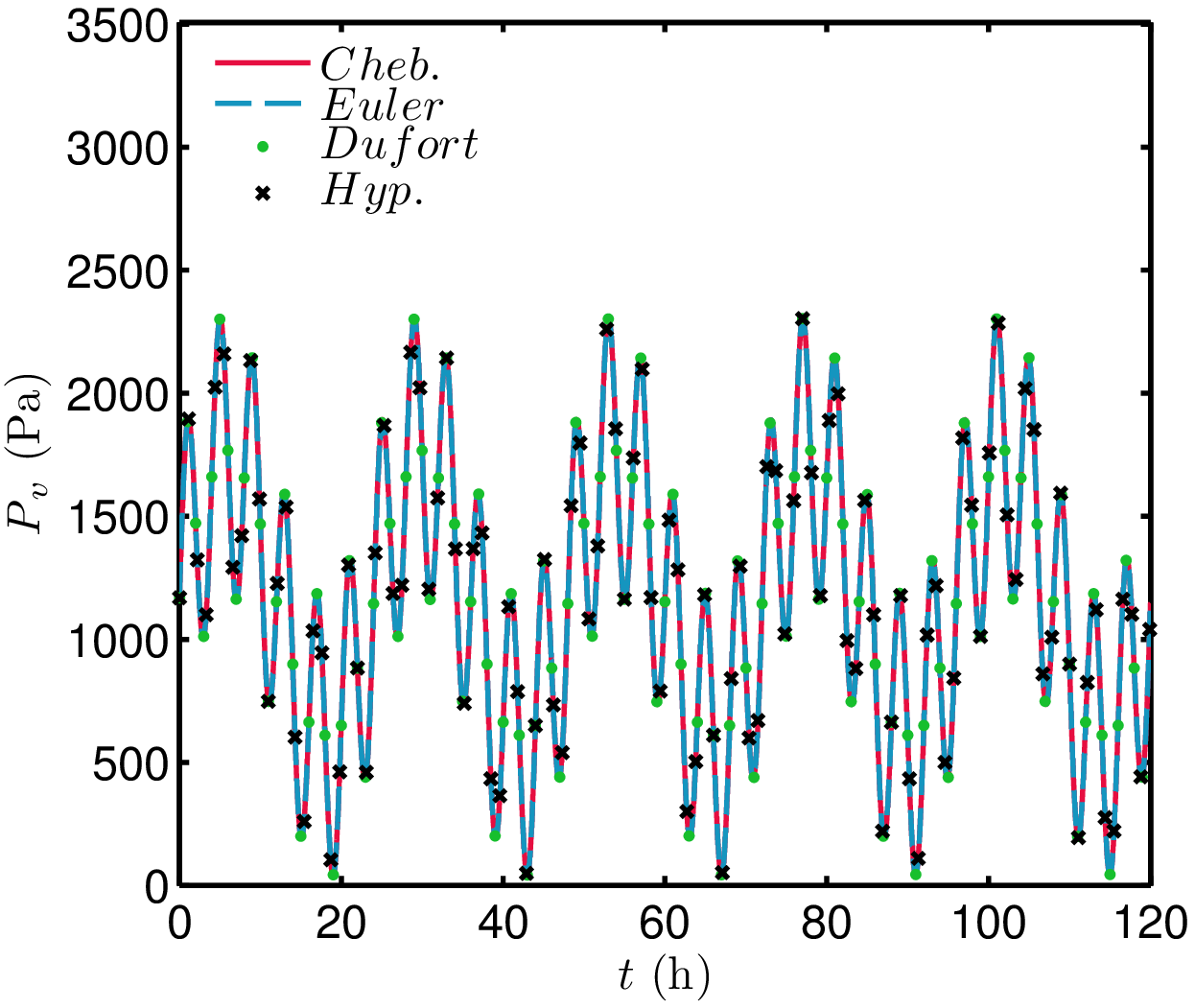}} \hspace{0.3cm}
  \subfigure[b][\label{fig_AN1:profil}]{\includegraphics[width=0.48\textwidth]{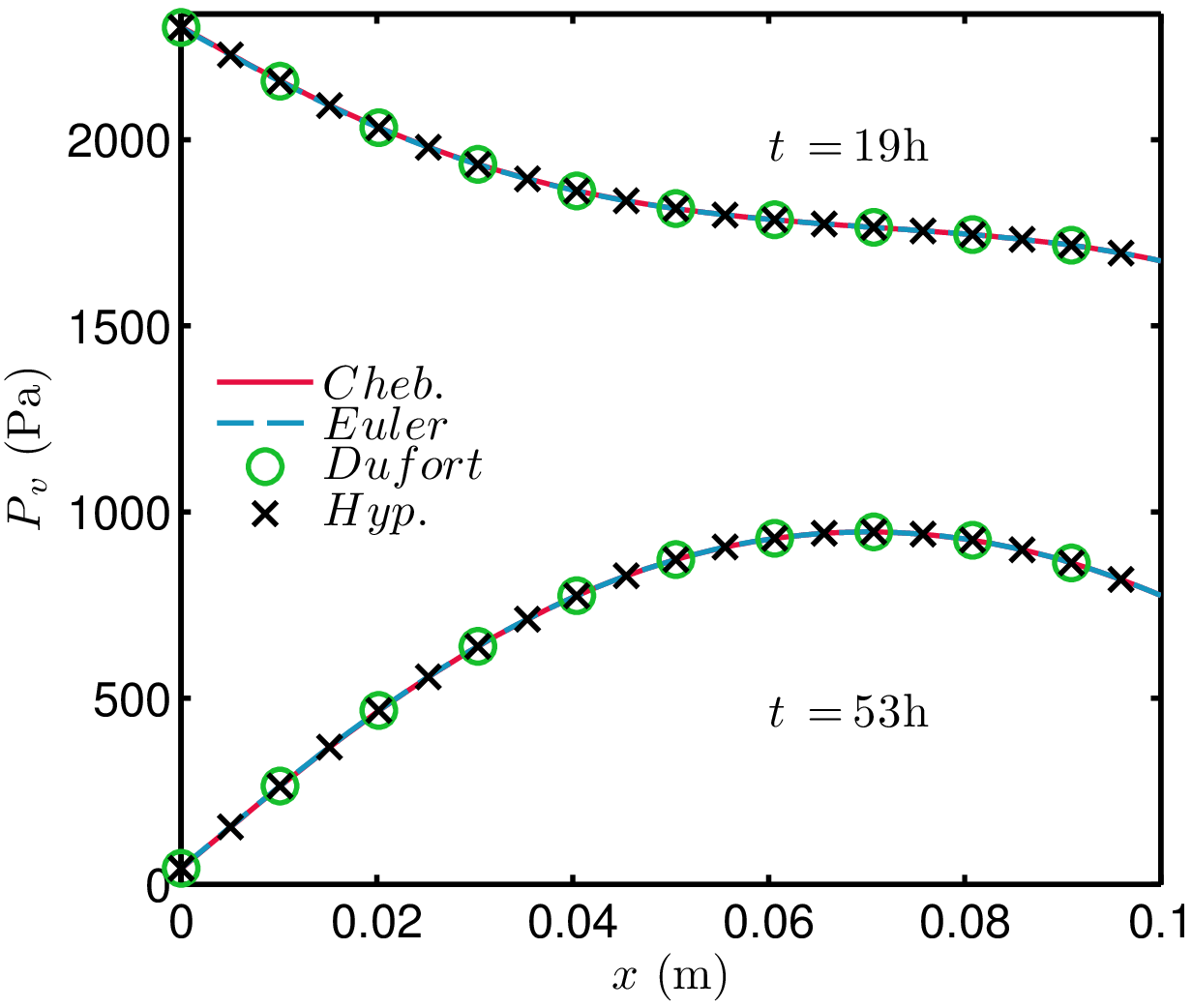}}
  \caption{\small\em Vapour pressure time evolution at $x \egal 0$  $\mathsf{m}$ (a) and profiles for $t \in \left\lbrace 19, \, 53\right\rbrace$  $\mathsf{h}$ (b).}
\end{figure}

\begin{figure}
  \centering
  \includegraphics[width=0.56\textwidth]{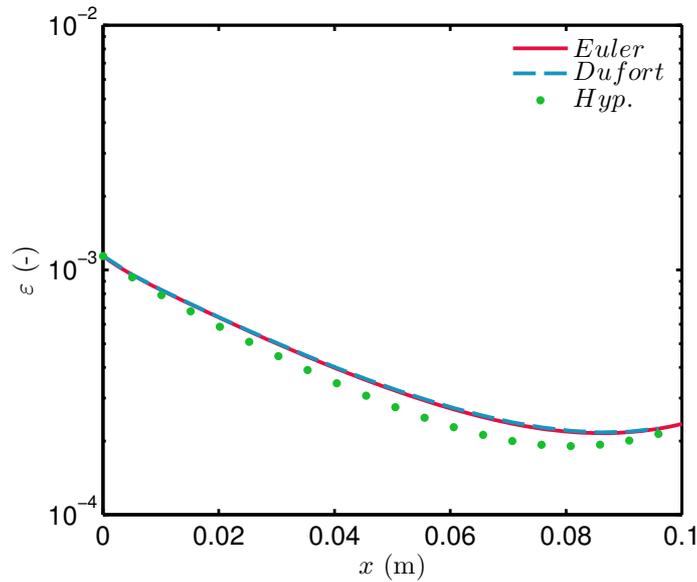}
  \caption{\small\em $\mathcal{L}_{\,2}$ error for fixed $\Delta t \egal 10^{\,-4}$.}
  \label{fig_AN1:err_fx}
\end{figure}

The solution of the problem has been first computed for a discretisation $\Delta x \egal 10^{-2}$ and $\Delta t \egal 10^{-4}$, respecting the CFL condition $\Delta t\, \leqslant\, 4.3 \dix{-4}$. For the hyperbolisation scheme, $\tau \egal \Delta t$. The physical phenomena are thus well represented, as illustrated in Figure~\ref{fig_AN1:x0time} with the time evolution of the vapour pressure at $x \egal 0$. The variations follow the ones of the left boundary conditions. It can be noted a good agreement between the four numerical schemes. Furthermore, the vapour pressure profile is shown in Figure~\ref{fig_AN1:profil} for $t \egal 19$  $\mathsf{h}$ and  $t \egal 52$  $\mathsf{h \,}$, corresponding to the highest and lowest vapour pressure values. All the numerical methods give accurate results as illustrated with the $\mathcal{L}_2$ error calculated as a function of $x$ in Figure~\ref{fig_AN1:err_fx}. The error for the hyperbolisation method is lower than the others. Indeed the hyperbolisation numerical scheme is of the order $\O (\Delta t^2)$ and the numerical solution is therefore more accurate.

\begin{figure}
  \centering
  \subfigure[a][\label{fig_AN1:err_fdt}]{\includegraphics[width=0.48\textwidth]{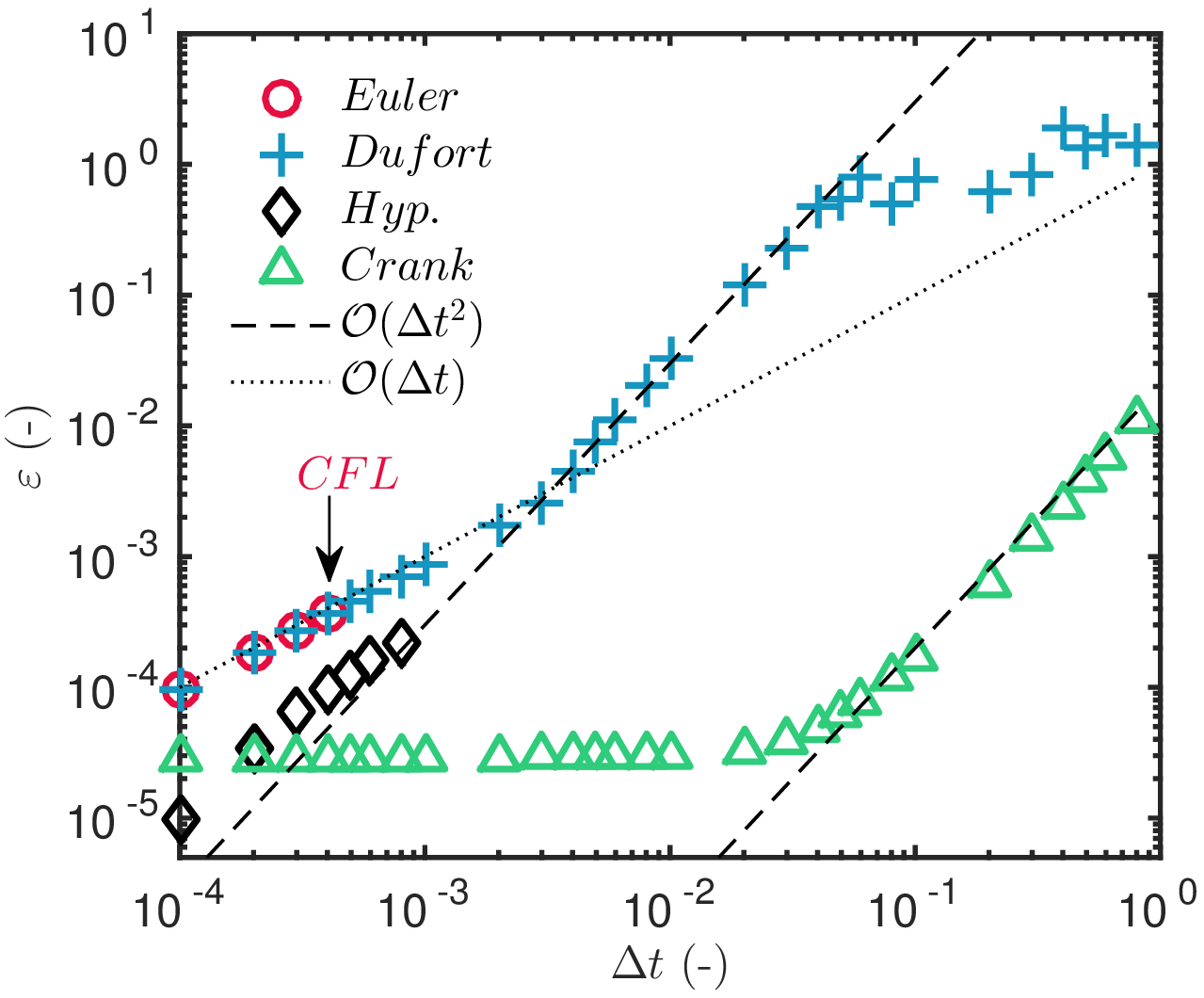}}
  \subfigure[b][\label{fig_AN1:errH_fdt}]{\includegraphics[width=0.48\textwidth]{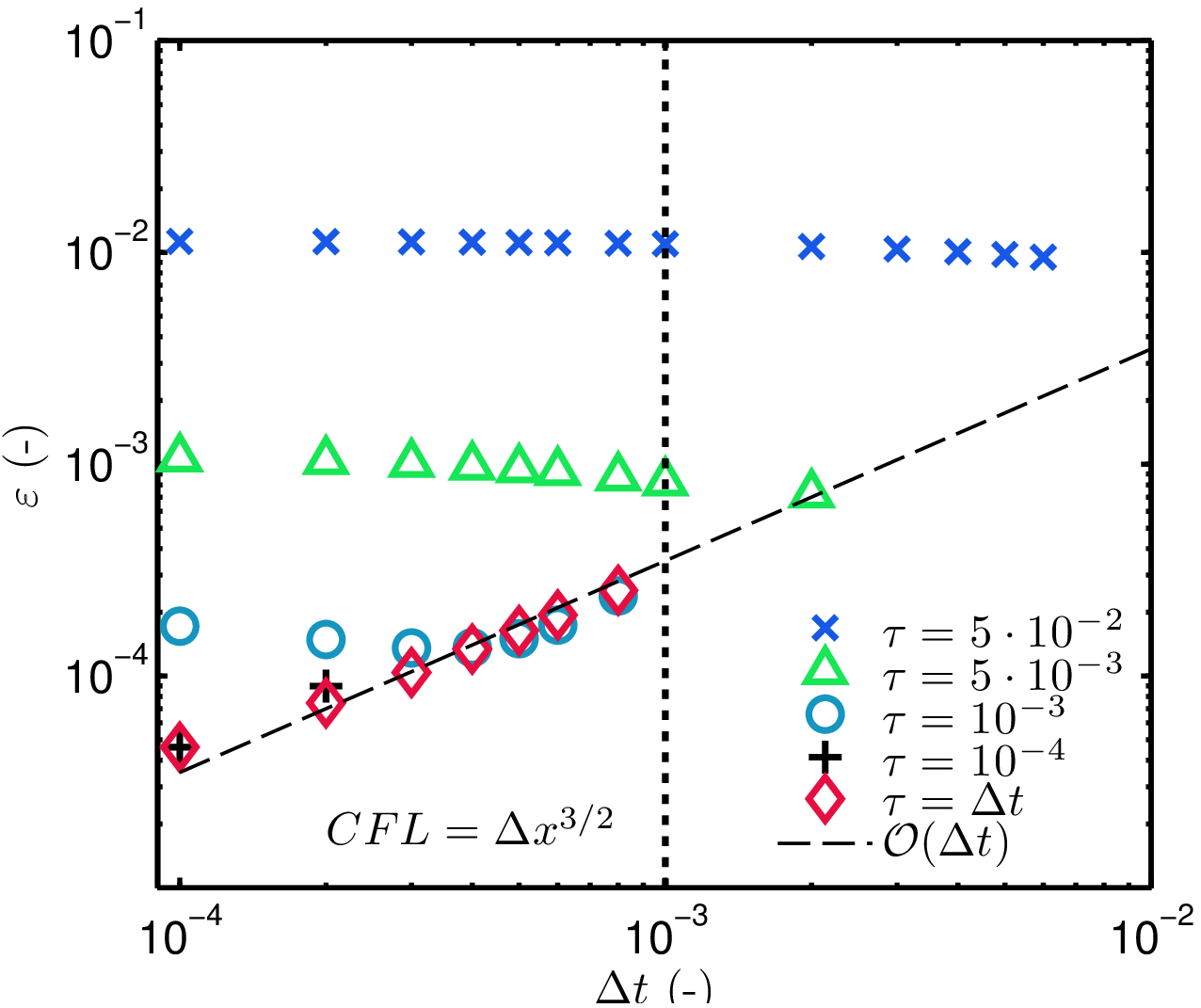}}
  \caption{\small\em $\mathcal{L}_{\,2}$ error as a function of $\Delta t$ for the \Eu, \DF ~and hyperbolisation schemes ($\Delta x \egal 10^{-2}$, $\tau \egal \Delta t$) (a) and for the hyperbolisation scheme ($\Delta x \egal 10^{-2}$).}
\end{figure}

A numerical analysis of the behaviour of the four numerical schemes has been carried out for different values of the temporal discretisation $\Delta t$. The spatial discretisation is maintained to $\Delta x \egal 10^{-2}$ and $\tau \egal \Delta t$ for the hyperbolisation scheme. The reference solution has been computed using the \texttt{Matlab} open source package \emph{Chebfun} \cite{Driscoll2014}. Using the function $\mathrm{pde23t}$, it enables to compute a numerical solution of a partial derivative equation using the \textsc{Chebyshev} functions. Results of the $\mathcal{L}_{\,2}$ error $\varepsilon$ are shown in Figure~\ref{fig_AN1:err_fdt}. Before the CFL limit, the errors of the \Eu, \DF ~and hyperbolisation schemes are proportional to $\O(\Delta t)$. As expected, the \Eu ~scheme enables to compute the solution as far the CFL condition is respected. Above this limit, the solution diverges. After the CFL limit, as unconditionally stable, the \DF ~scheme computes the solution. An interesting point is that the error is proportional to $\O(\Delta t^2)$. The error of the \DF ~scheme computed solution becomes too high for $\Delta t \, \geqslant \, 3 \dix{-2}$. For this case, $\Delta t\ \propto\ \Delta x^{\,\frac{3}{2}}$, therefore, the \DF ~scheme is first order accurate in space $\O(\Delta x)$, explaining why the error is lower for the hyperbolisation scheme.

As the \Eu ~scheme, the hyperbolisation scheme has a stability condition to respect as reported in Eq.~\eqref{eq:cond_hyp}. For the case $\tau \egal \Delta t$, it corresponds to $\Delta t \ \leqslant \ 9 \dix{-4} \, $. This limit is higher than the CFL condition for the \Eu ~scheme. The error of the hyperbolisation scheme varies with the choice of the parameter $\tau$. Figure~\ref{fig_AN1:errH_fdt} gives the variation of the error $\varepsilon$, for the hyperbolisation scheme, as a function of $\Delta t$, for different values of parameter $\tau$. The error $\varepsilon$ reaches a limit lower than the parameter $\tau \,$. It can be verified that for the choice $\tau \egal \nu \, \Delta x \, $, the stability condition corresponds to $\Delta x^{\,\frac{3}{2}} \,$. For this case, the choice $\tau \ \leqslant \ \Delta t$ permits to compute the solution with the best accuracy.

%%% ------------------------------------------------------------------------ %%%

\section{Extension for nonlinear transfer}

The previous case study investigated the use of three numerical schemes for computing the solution of a linear problem of moisture diffusion. This second case study considers now nonlinear diffusion, due to material properties depending on the moisture content $\dms\left(\, u \, \right)$ and  $\cms\left(\, u \, \right)$. This case will be investigated via the \DF ~and the improved \CN ~schemes. The hyperbolisation approach is not considered, as a stability condition was observed previously in the linear case. First, the \DF ~and \CN ~schemes are detailed for the nonlinear case. For this, Eq.~\eqref{eq:moisture_dimensionlesspb_1D} is re-called with a simplified notation:
\begin{align}\label{eq:heat_NL}
  & c(u) \, \pd{u}{t} \egal \pd{}{x} \Biggl[ \, d(u) \, \pd{u}{x}  \, \Biggr] \,.
\end{align}

%%% ------------------------------------------------------------------------ %%%

\subsection{The \mCN ~scheme}

The straightforward application of the \CN ~scheme to Eq. \eqref{eq:heat_NL} yields the following scheme:
\begin{align}\label{eq:CN_NL}
  c_{\,j}^{\,n} \, \frac{u_{\,j}^{\,n+1}\ -\ u_{\,j}^{\,n}}{\Delta t} \egal \frac{1}{\Delta x} \Biggl[\, \left(\, d \;\pd{u}{x} \, \right)_{\,j+\half}^{\,n+\half}   \moins \left(\, d \;\pd{u}{x} \, \right)_{\,j-\half}^{\,n+\half}   \, \Biggr] \,,
\end{align}
with 
\begin{align*}
  \left(\, d\pd{u}{x} \, \right)_{\,j+\half}^{\,n+\half} & \egal \frac{1}{2} \Biggl[\, \left(\, d \; \pd{u}{x} \, \right)_{\,j+\half}^{\,n+1} \plus \left(\, d \; \pd{u}{x} \, \right)_{\,j+\half}^{\,n}\,\Biggr] \\
  & \egal 
  \frac{1}{2 \ \Delta x} \Biggl[\, d_{\,j+\half}^{\,n+1} \, \Big(\, u_{\,j+1}^{\,n+1} \moins u_{\,j}^{\,n+1} \,\Big) \plus d_{\,j+\half}^{\,n} \, \Big(\, u_{\,j+1}^{\,n} \moins u_{\,j}^{\,n} \,\Big) \,\Biggr]\,.
\end{align*}
However, this approach leads to deal with nonlinearities due to the evaluation of quantities (as $d_{\,j+\half}^{\,n+1}$) at the upcoming time layer $t \egal t^{\,n+1}$. To deal with this issue, linearisation techniques as \textsc{Picard} or \textsc{Newton}--\textsc{Raphson} ones \cite{Raphson1690, Cajori1911}, requiring a high number of sub-iterations. To overcome these difficulties, it is possible to evaluate the diffusion coefficient at the actual time layer instead of the upcoming \cite{Ascher1995}. Thus, the diffusion flux at the interface  becomes:
\begin{align*}
  \left(\, d \; \pd{u}{x} \, \right)_{\,j+\half}^{\,n+\half} \egal \frac{1}{2 \ \Delta x} \Biggl[\, d_{\,j+\half}^{\,n} \, \Big(\, u_{\,j+1}^{\,n+1} \moins u_{\,j}^{\,n+1} \,\Big) \plus d_{\,j+\half}^{\,n} \, \Big(\, u_{\,j+1}^{\,n} \moins u_{\,j}^{\,n} \,\Big) \,\Biggr] \,.
\end{align*}

Finally, the \mCN ~schemes yields to: 
\begin{align*}
  & \Biggl[ \, 1 \plus \frac{\Delta t}{2 \ \Delta x^2} \, \left(\, d_{\,j+\half}^{\,n} \plus d_{\,j-\half}^{\,n}\,\right) \, \Biggr] \, u_{\,j}^{\,n+1} \moins \frac{\Delta t}{2 \ \Delta x^2} \, d_{\,j+\half}^{\,n} \, u_{\,j+1}^{\,n+1} \moins \frac{\Delta t}{2 \ \Delta x^2} \, d_{\,j-\half}^{\,n} \, u_{\,j-1}^{\,n+1} \\
  & \egal \Biggl[ \, 1 \moins \frac{\Delta t}{2 \ \Delta x^2} \, \left(\, d_{\,j+\half}^{\,n} \plus d_{\,j-\half}^{\,n}\,\right) \, \Biggr] \, u_{\,j}^{\,n} \plus \frac{\Delta t}{2 \ \Delta x^2} \, d_{\,j+\half}^{\,n} \, u_{\,j+1}^{\,n} \plus \frac{\Delta t}{2 \ \Delta x^2} \, d_{\,j-\half}^{\,n} \, u_{\,j-1}^{\,n} \,.
\end{align*}
The combination of implicit-explicit (\emph{IMEX)} approaches clearly appear in this formulation. The major advantage over the classical \CN ~scheme is to avoid sub-iterations in the solution procedure, without loosing the accuracy and the stability.

%%% ------------------------------------------------------------------------ %%%

\subsection{The \DF ~scheme}

In the nonlinear case, the \DF ~numerical schemes is written as: 
\begin{align}\label{eq:DF_NL}
  c_{\,j}^{\,n} \, \frac{u_{\,j}^{\,n+1}\ -\ u_{\,j}^{\,n-1}}{2\,\Delta t} \egal \frac{1}{\Delta x} \Biggl[\, \left(\, d\pd{u}{x} \, \right)_{\,j+\half}^{\,n}   \moins \left(\, d\pd{u}{x} \, \right)_{\,j-\half}^{\,n}   \, \Biggr] \,.
\end{align} 
The right-hand side term can be expressed as: 
\begin{align}\label{eq:DF_NL_RHS}
  & \frac{1}{\Delta x} \left(\, \left(\, d\pd{u}{x} \, \right)_{\,j+\half}^{\,n}   \moins \left(\, d\pd{u}{x} \, \right)_{\,j-\half}^{\,n}   \, \right) \egal \frac{1}{\Delta x^2} \left(\, \, d_{\,j+\half}^{\,n}  \, u_{\,j+1}^{\,n} \plus \, d_{\,j-\half}^{\,n}  \, u_{\,j-1}^{n} \moins \left(\, d_{\,j+\half}^{\,n}   \plus  d_{\,j-\half}^{\,n}  \, \right) u_{\,j}^{\,n} \,\right) \,.
\end{align}
Using the \DF ~stencil (see Figure~\ref{fig:stencil_Dufort}), the term $u_{\,j}^{\,n}$ is replaced by $\dfrac{u_{\,j}^{\,n+1} \plus u_{\,j}^{\,n-1}}{2}\,$. Thus, considering Eq.~\eqref{eq:DF_NL}, the \DF ~schemes can be expressed as an explicit scheme:
\begin{align*}
  & u_{\,j}^{\,n+1} \egal \frac{\lambda_{\,1}}{\lambda_{\,0} \plus \lambda_{\,3}} \cdot u_{\,j+1}^{\,n} \plus \frac{\lambda_{\,2}}{\lambda_{\,0} \plus \lambda_{\,3}} \cdot u_{\,j-1}^{\,n} \plus \frac{\lambda_{\,0} \moins \lambda_{\,3}}{\lambda_{\,0} \plus \lambda_{\,3}} \cdot u_{\,j}^{\,n-1} \,,
\end{align*}
with 
\begin{align*}
  & \lambda_{\,0} \ \eqdef \ c_{\,j}^{\,n} \,, && \lambda_{\,1} \ \eqdef \  \frac{2 \, \Delta t}{\Delta x^2} \; d_{j+\half}^{\,n} \,, \\
  & \lambda_{\,2} \ \eqdef \  \frac{2 \, \Delta t}{\Delta x^2} \; d_{j-\half}^{\,n} \,,&& \lambda_{\,3} \ \eqdef \ \frac{\Delta t}{\Delta x^2} \, \left(\, d_{j+\half}^{\,n} + d_{j-\half}^{\,n} \, \right) \,.
\end{align*}
When dealing with the nonlinearities of the material properties, an interesting feature of explicit schemes is that it does not require any sub-iterations (using \textsc{Newton}--\textsc{Raphson} approach for instance). At the time iteration $n$, the material properties $c_{\,j}$, $d_{j+\half}$, $d_{j-\half}$ are \emph{explicitly} calculated at $t^{\,n}$. It should be noted that the material properties evaluated at $j \plus \half$ is formulated as:     
\begin{align*}      
  d_{j+\half} \egal d \biggl(\, \frac{u_{j} \plus u_{\j+1}}{2} \,\biggr) \,.      
\end{align*}      

%%% ------------------------------------------------------------------------ %%%

\subsection{Numerical application}
\label{sec:case_nonlinear1}

\begin{figure}
  \centering
  \subfigure[a][\label{fig_AN2:cM}]{\includegraphics[width=0.45\textwidth]{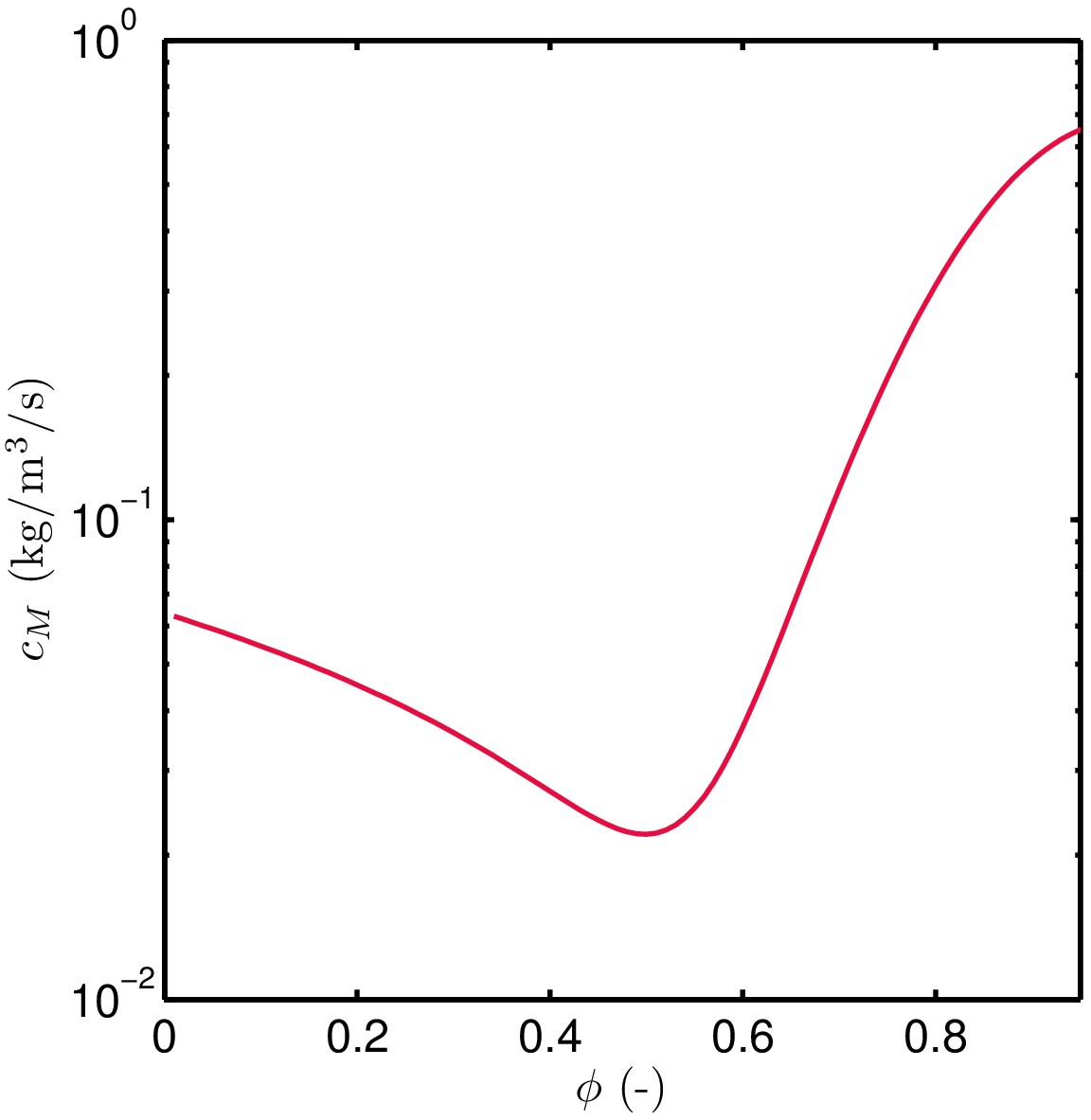}}
  \subfigure[b][\label{fig_AN2:dM}]{\includegraphics[width=0.48\textwidth]{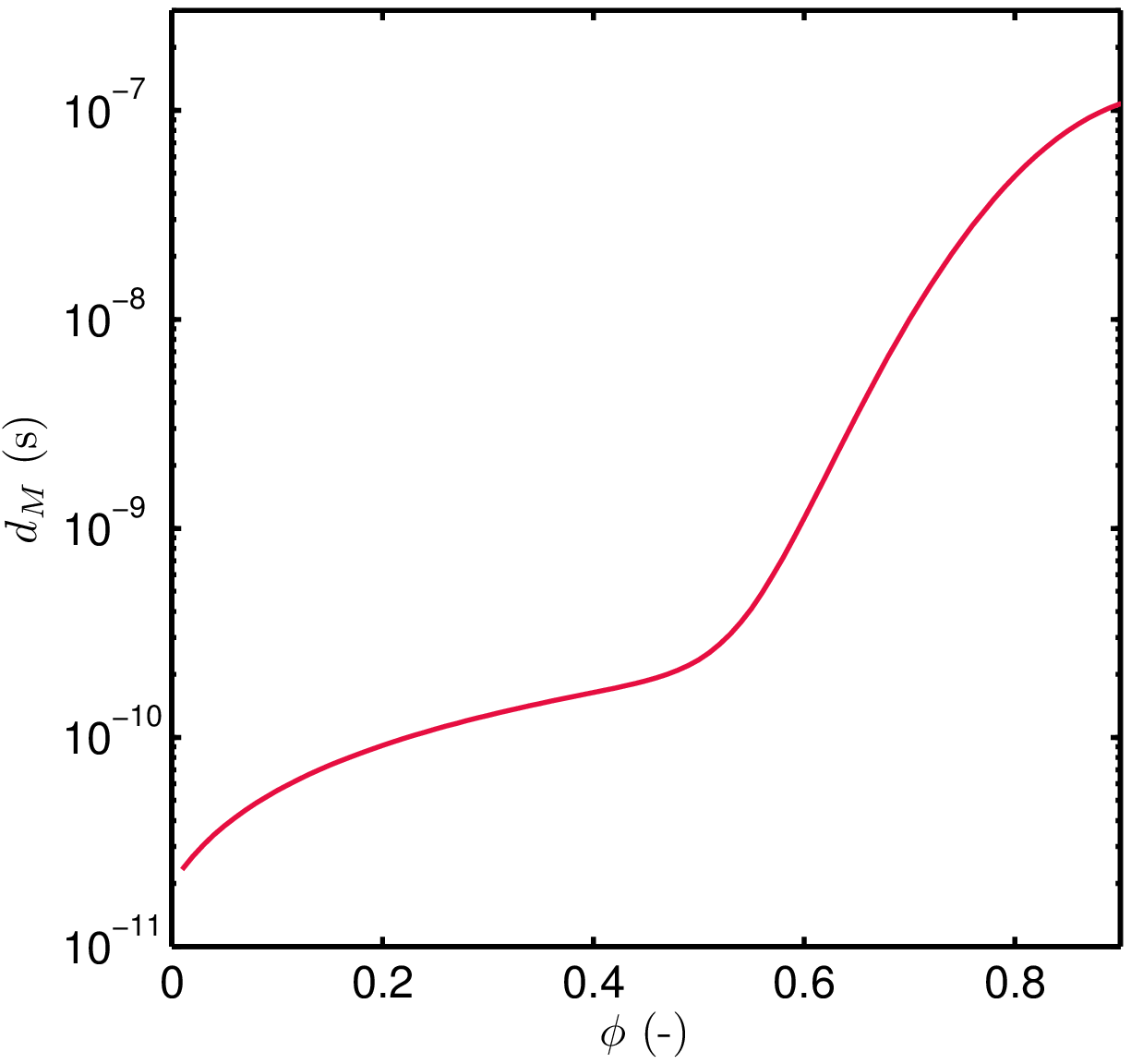}}
  \caption{\small\em Variation of the moisture storage $\cm$ (a) and diffusion $\dm$ (b) as a function of the relative humidity $\phi\,$.}
\end{figure}

\begin{figure}
  \centering
  \includegraphics[width=0.59\textwidth]{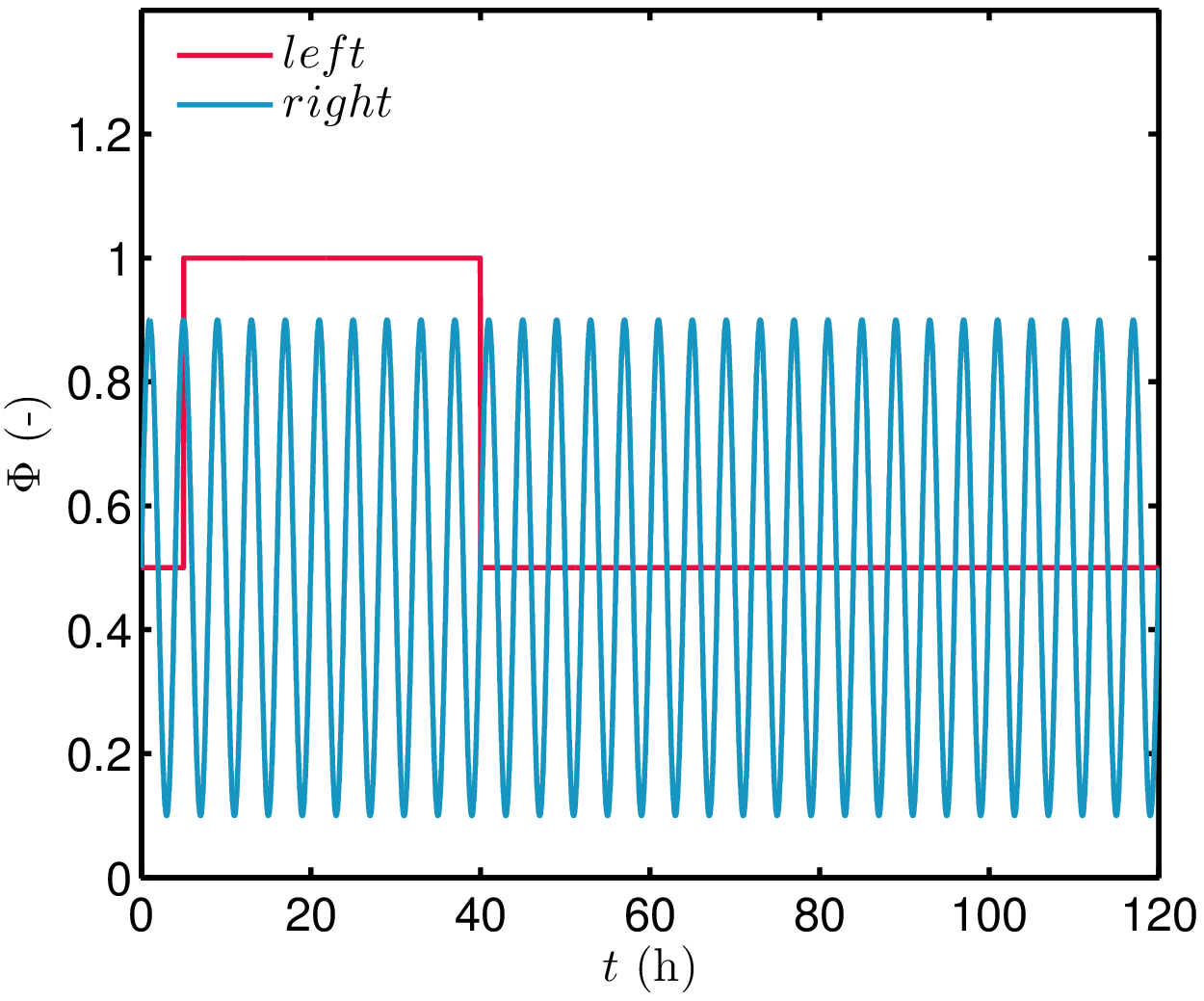}
  \caption{\small\em Boundary conditions.}
  \label{fig_AN2:BC}
\end{figure}

From a physical point of view, the storage and diffusion coefficients are given in Figures \ref{fig_AN2:cM} and \ref{fig_AN2:dM}. Their variations with the relative humidity are similar to the load bearing material from \cite{Janssen2014}. The initial vapour pressure is uniform $\Pvi \egal 1.16 \dix{3}$  $\mathsf{Pa}$. No moisture flow is taken into account at the boundaries. The ambient vapour pressure at the boundaries are illustrated in Figure~\ref{fig_AN2:BC}. At the left boundary, it has a fast drop until the saturation state and at the right boundary, it has a sinusoidal variation. The material is thus excited until the capillary state. The convective vapour transfer coefficients are set to $2 \cdot 10^{-7}$  $\mathsf{s/m}$ and $3 \cdot 10^{-8}$  $\mathsf{s/m}$ for the left and right boundary conditions, respectively. The final simulation time is also fixed to $120$ hours. As in the previous case study, the dimensionless values can be found in Appendix~\ref{annexe:dimensionless}.

\begin{figure}
  \centering
  \subfigure[a][\label{fig_AN2:time_Pv}]{\includegraphics[width=0.48\textwidth]{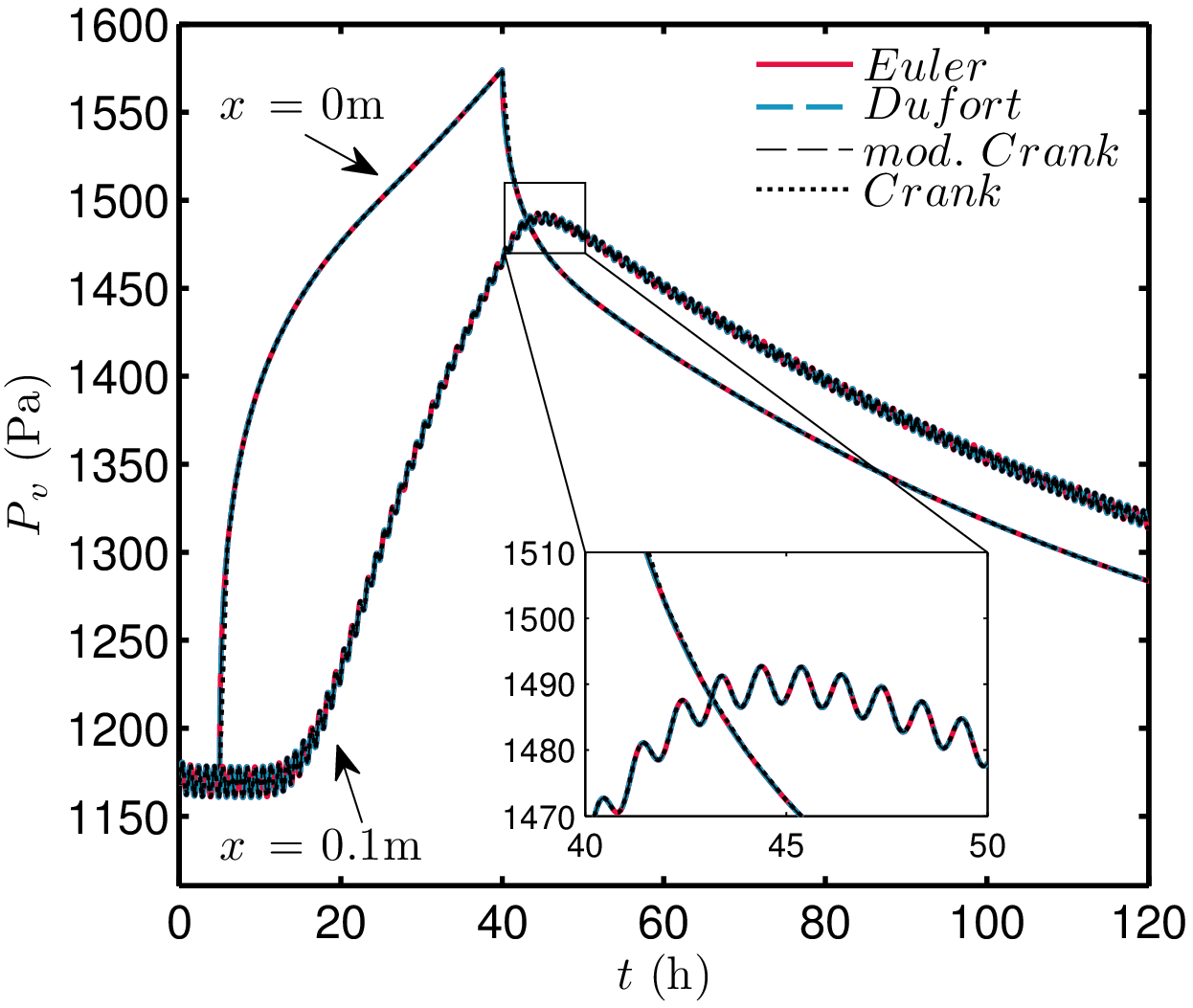}}
  \subfigure[b][\label{fig_AN2:profil_Pv}]{\includegraphics[width=0.48\textwidth]{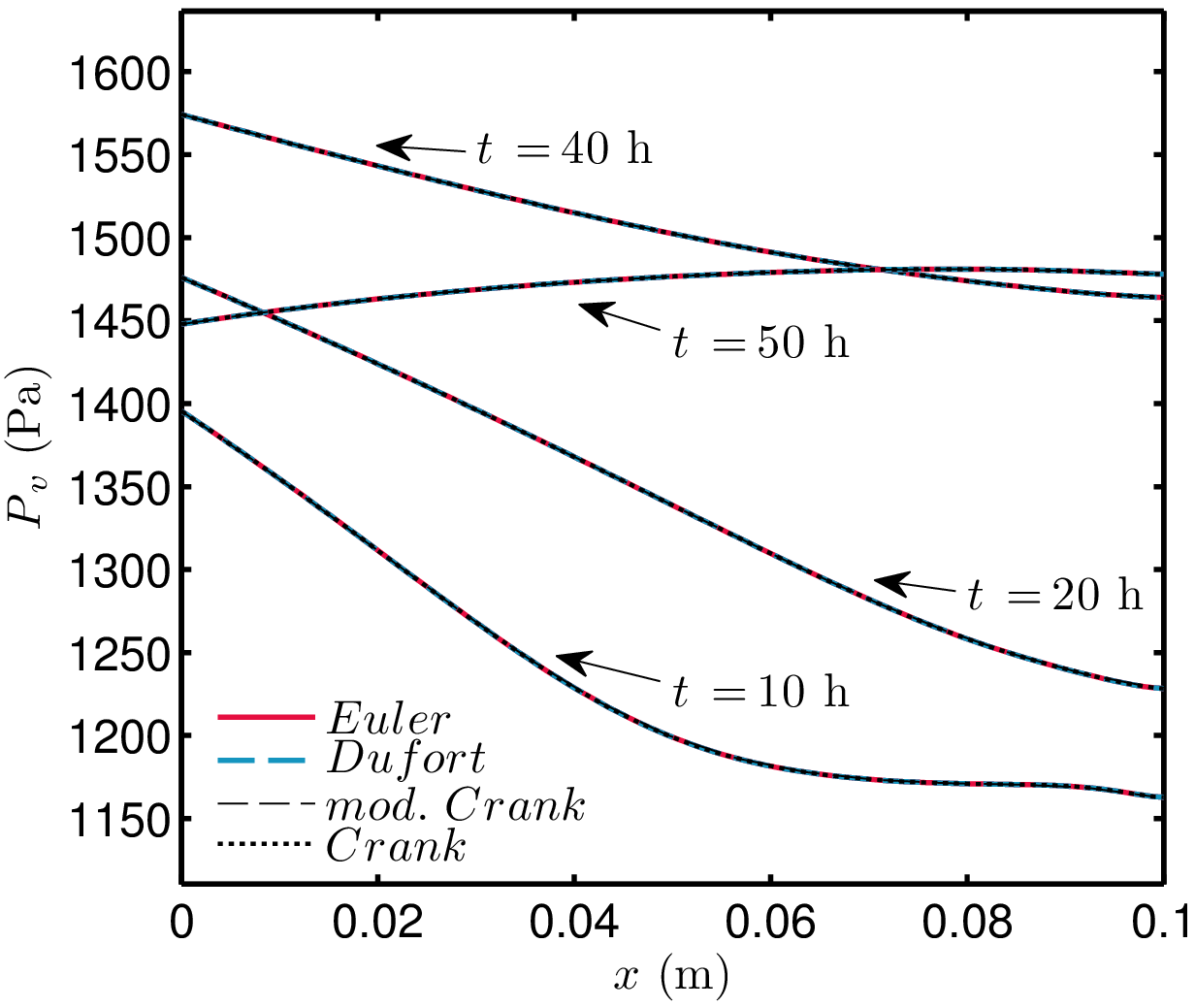}}
  \caption{\small\em Vapour pressure time evolution at $x \in \left\lbrace 0, \, 0.1 \right\rbrace$  ~$\mathsf{m}$ (a) and profiles for $t \in \left\lbrace 10, \, 20, \, 40, \, 50 \right\rbrace$  $\mathsf{h}$ (b).}
\end{figure}

\begin{figure}
  \centering
  \includegraphics[width=0.59\textwidth]{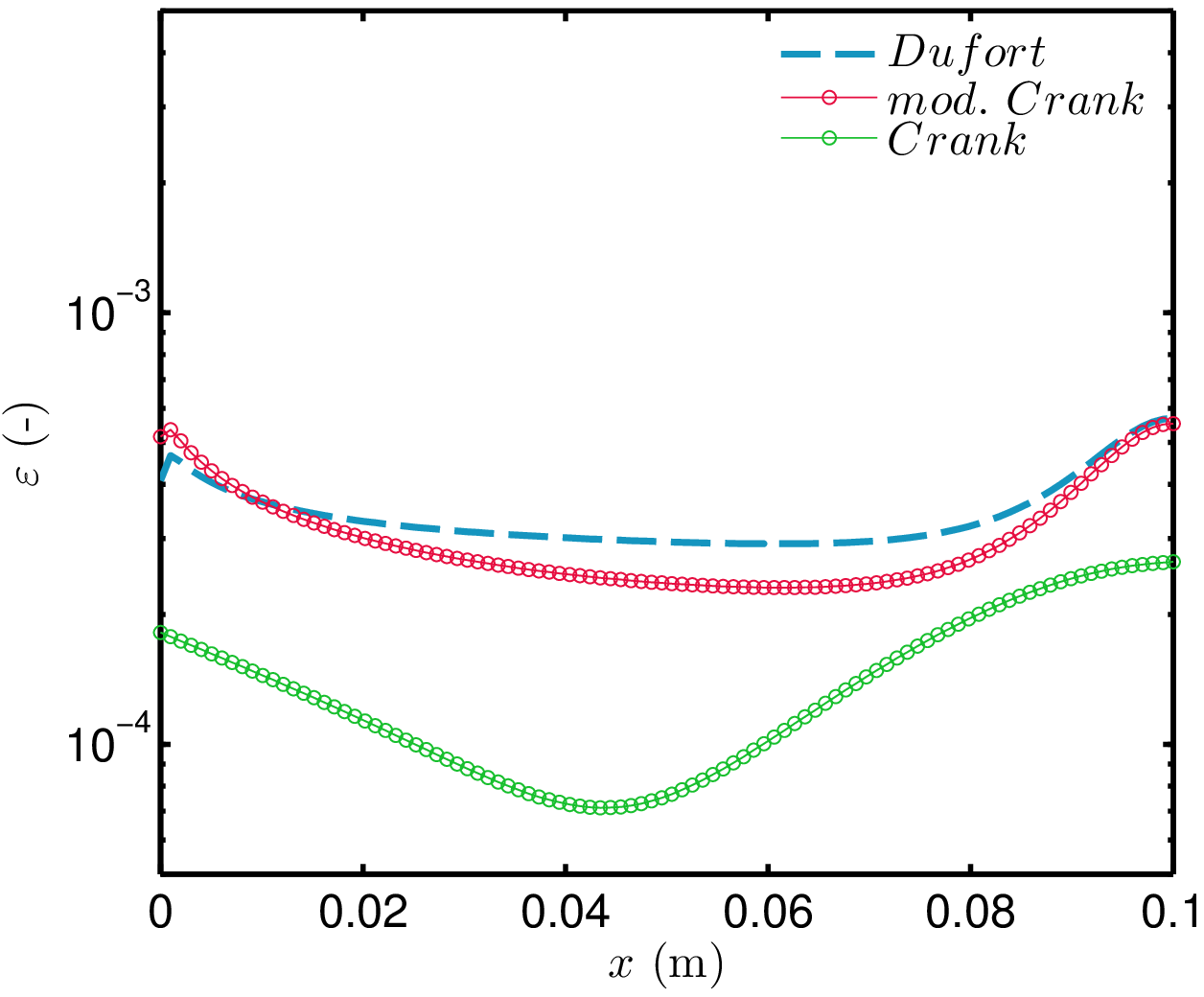}
  \caption{\small\em $\mathcal{L}_{\,2}$ error for a fixed $\Delta t \egal 6 \dix{-6}\,$.}
  \label{fig_AN2:err_fx}
\end{figure}

The solution of the problem has been computed with the following discretisation parameters: $\Delta t \egal 10^{\,-4}$ and $\Delta x \egal 10^{\,-2}$. For this, the \DF, the \mCN ~and the standard \CN ~numerical schemes have been used. A sufficiently converged solution, computed with a \textsc{Euler} explicit scheme, is taken as reference. For the latter, the tolerance is set to $\epsilon \, \leqslant \, 0.01 \cdot \Delta t^2$ to ensure the convergence of the sub-iterations. The time variation of the vapour pressure according for the bounding points is given in Figure~\ref{fig_AN2:time_Pv}. The vapour pressure in the material is increasing according to the variation at the left boundary condition. There is a delay between the vapour pressure at the left ($x \egal 0$  $\mathsf{m}$) and right ($x \egal 0.1$ $\mathsf{m}$) bounding points. This increase can also be observed on the four profiles of vapour pressure illustrated in Figure~\ref{fig_AN2:profil_Pv}. Furthermore, a break in the slope of the increase of the vapour pressure can be noticed at $t \egal 12$  $\mathsf{h \,}$, due to the nonlinear behaviour of the material. The vapour pressure at $x \egal 0.1$  $\mathsf{m}$ slowly oscillated according to the right boundary condition. All the solutions, computed with each different numerical schemes, have good agreement to represent the physical phenomena and their $\mathcal{L}_{\,2}$ error with the reference solution is lower than $10^{\,-3}$, as reported in Figure~\ref{fig_AN2:err_fx}.

\begin{figure}
  \centering
  \includegraphics[width=0.59\textwidth]{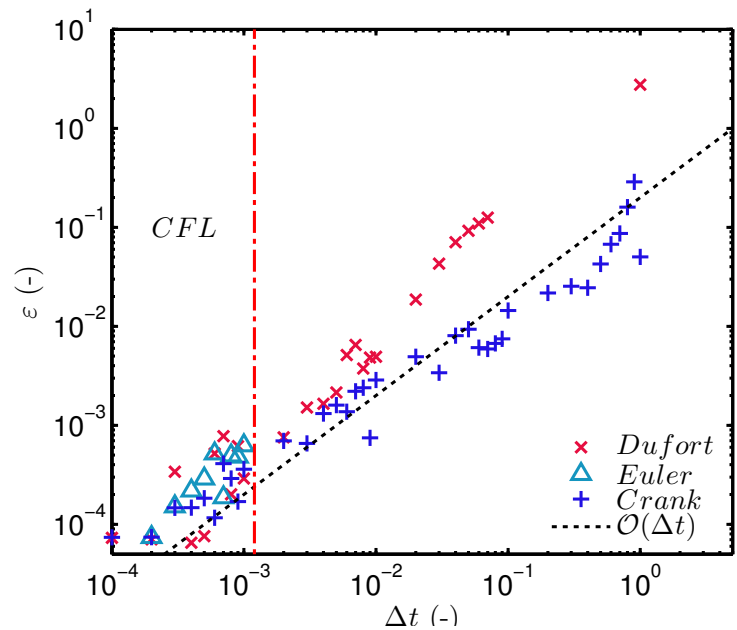}
  \caption{\small\em $\mathcal{L}_{\,2}$ error (a) as a function of $\Delta t\,$.}
  \label{fig_AN2:err_f_dt}
\end{figure}

The solution has been computed for different values of $\Delta t$, maintaining $\Delta x \egal 10^{\,-2}$. For each value of $\Delta t$, the $\mathcal{L}_2$ error has been computed between the numerical and a sufficiently converged reference solution. Results are given in Figure~\ref{fig_AN2:err_f_dt}. The equivalent CFL condition has been computed as:
\begin{align*}
  \Delta t\ \leqslant\ \frac{\Delta x^2}{2 \, \, \max \left( \frac{\dms\left(\, u \,\right)}{\cms\left(\, u \,\right)} \right)}\,.
\end{align*}

As for the linear case, the \Eu ~explicit scheme enables to compute the solution while the CFL condition is respected. The \DF ~and \mCN ~numerical schemes are unconditionally stable. An interesting observation is that the error of the \DF ~and \mCN ~schemes are proportional to $\O(\Delta t)$. The modification of the \CN ~scheme, in order to avoid the sub-iterations due to nonlinearities, loose the $\O(\Delta t^2)$ accuracy.

\begin{figure}
  \centering
  \subfigure[a][\label{fig_AN2:time_Pv_bad_dt_0}]{\includegraphics[width=0.48\textwidth]{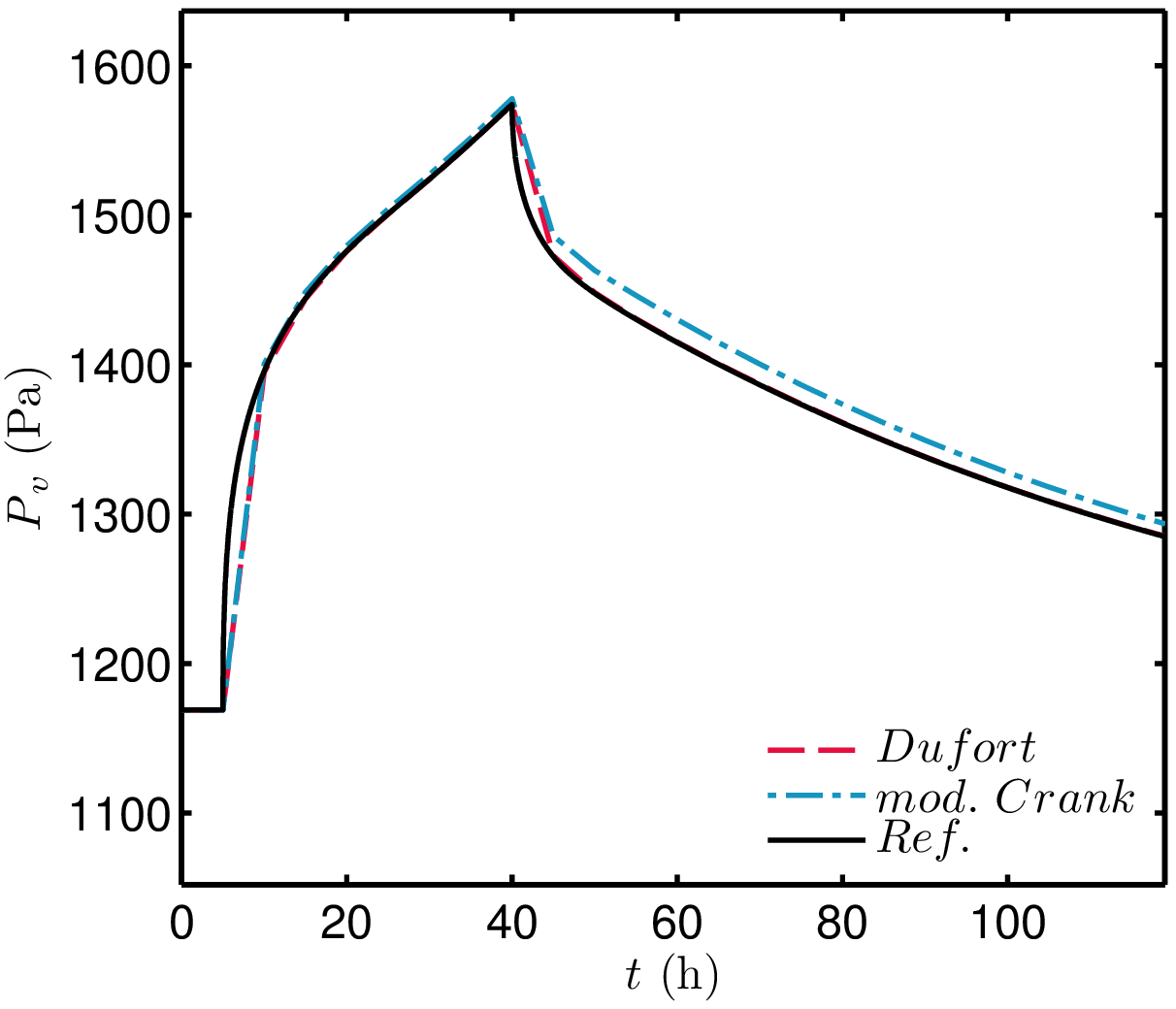}}
  \subfigure[b][\label{fig_AN2:time_Pv_bad_dt_1}]{\includegraphics[width=0.48\textwidth]{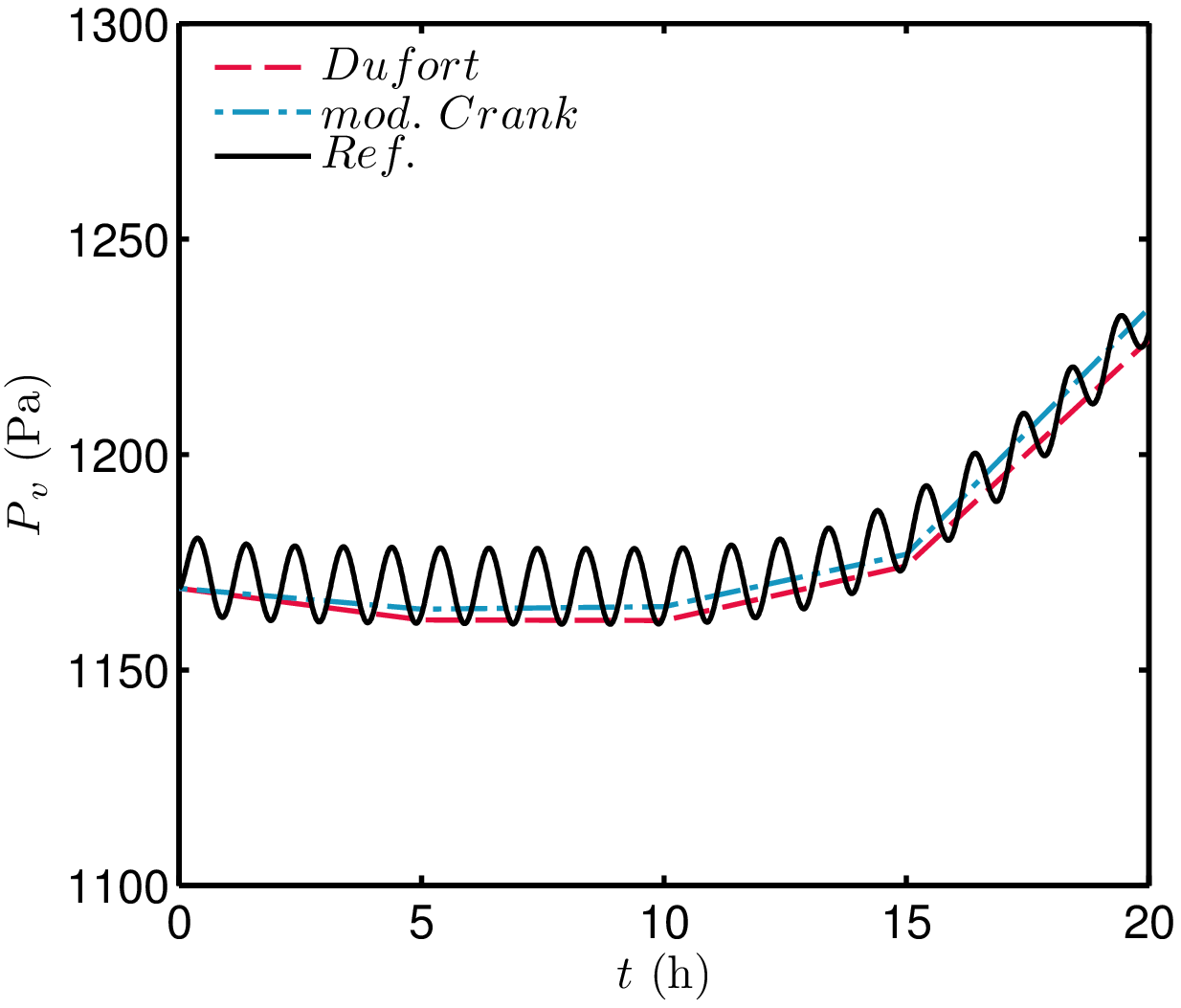}}
  \caption{\small\em Comparison of reference solution and the one computed with \DF ~and \mCN ~schemes for $\Delta t \egal 10^{\,-1}$ at $x \egal 0$ $\, \mathsf{m}$ (a) and $x \egal 0.1$ $\, \mathsf{m}$ (b).}
\end{figure}

\begin{figure}
  \centering
  \includegraphics[width=0.59\textwidth]{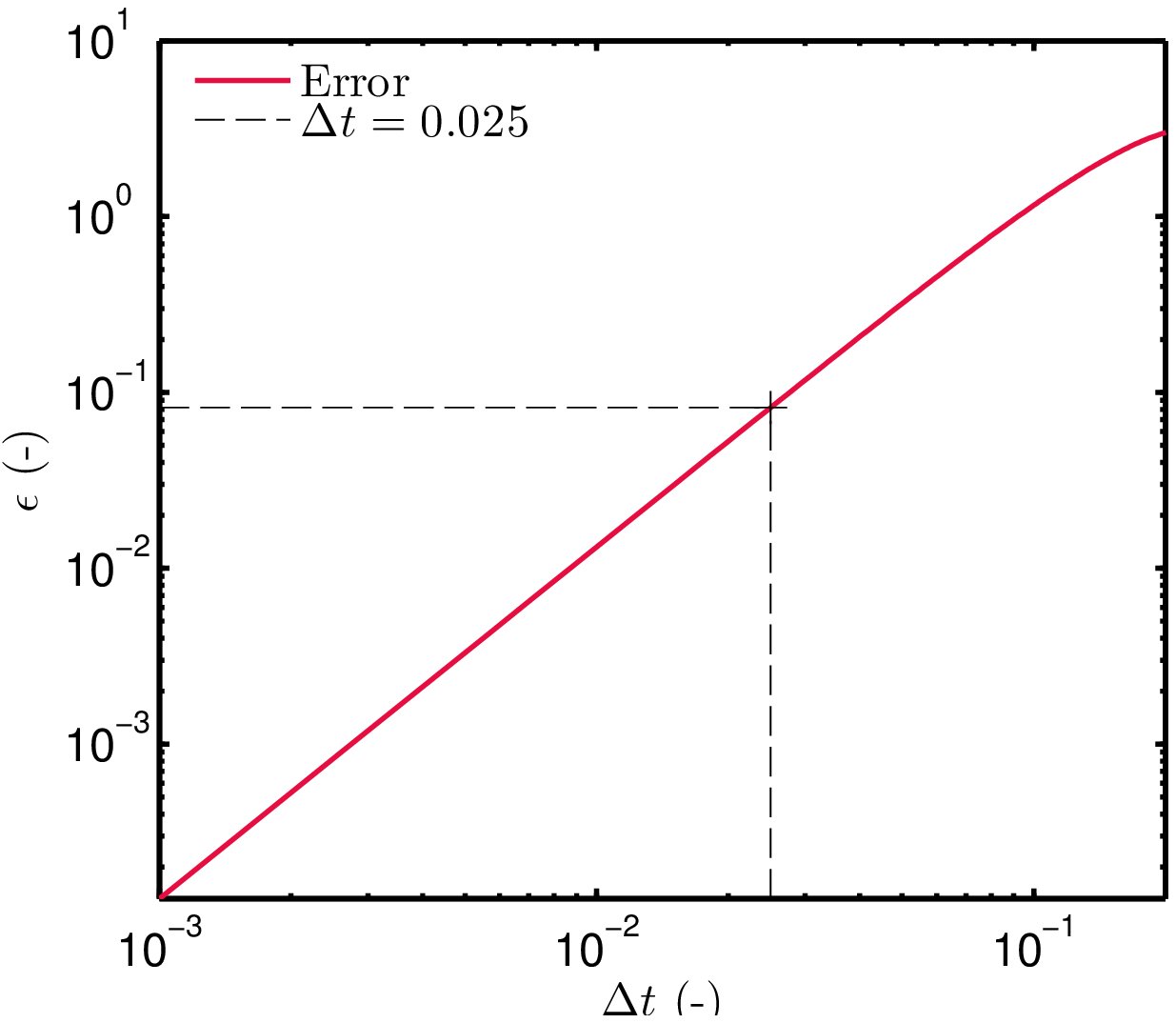}
  \caption{\small\em Error due to time discretisation for the right boundary condition $\uR$ ($A \egal 0.8$ and $\omega \egal 2 \pi$).}
  \label{fig_AN2:err_BC}
\end{figure}

Even with $\Delta t$ increasing, the schemes are able to compute a solution. However, the choice of the time discretisation is an important issue to represent accurately the physical phenomenon. Figures~\ref{fig_AN2:time_Pv_bad_dt_0} and \ref{fig_AN2:time_Pv_bad_dt_1} show the vapour pressure evolution computed with $\Delta t \egal 10^{\,-1}$. The solution lacks of accuracy comparing to the reference solution. For instance, at $t \egal 40$, the solution does not represent accurately the decrease of the vapour pressure.  At $x \egal 0.1$  $\mathsf{m\, }$, $\Delta t$ is too large to follow the dynamic of the boundary condition. The error due to the time discretisation of a sinusoidal boundary condition can be expressed as:
\begin{align*}      
  \epsilon \egal \left| A \, \omega \, \cos \left(\, \omega n \Delta t \, \right)     
  \moins A \, \frac{\sin \left(\, \omega \left(\,n+1\,\right) \Delta t  \, \right) - \sin \left(\, \omega n \Delta t  \, \right)}{\Delta t} \right| \,,     
\end{align*}
where $A$ and $\omega$ are the amplitude and the frequency of the signal. The error $\epsilon$ is given in Figure~\ref{fig_AN2:err_BC} as a function of $\Delta t$ for this case study. If an accuracy of $\epsilon \, \leqslant \, 10^{\,-1}$ is required, a time discretisation lower than $\Delta t \, \leqslant \, 2 \dix{-2}$ is needed. Therefore, for an unconditionally stable scheme, the choice of the time discretisation depends on the variation of the boundary conditions, as well as the diffusion time in the material, in order to compute a solution representing accurately the physical phenomenon.

With $N$, the number points due to spatial discretisation, at each time iteration $t^{\,n}$ the numbers of operations for each schemes scales with: 
\begin{align*}
  & \text{\textsc{Euler} implicit:} && \O(N_{\,\mathrm{NL}} \cdot N) \,, \\
  & \text{\CN:} && \O(N_{\,\mathrm{NL}} \cdot 2 \cdot N) \,, \\
  & \text{\textsc{Euler} explicit:} && \O(N) \,, \\
  & \text{\DF} && \O(N) \,, \\
  & \text{\mCN:} && \O(2 \cdot N) \,. \\
\end{align*}
The standard \textsc{Euler} implicit and \CN ~schemes require $N_{\,\mathrm{NL}}$ sub-iterations to treat the nonlinearities  (using \textsc{Newton}--\textsc{Raphson} approach for instance). The \CN ~approach needs twice more operations as it combines explicit and implicit approaches. The number of operations of the explicit schemes, as \textsc{Euler} and \DF, scales with the direct computation of the solution $u^{\,n}\,$. The \mCN ~does not require sub-iterations as the coefficients are expressed at the current time layer. Thus, the number of operations required is $\O(2 \cdot N)$. Generally $\O(N_{\,\mathrm{NL}}) \, \gg \, 1$, as reported in \cite{Janssen2014} where the number of sub-iterations scales between $10$ to $30\,$. Therefore, for the same discretisation, the \CN ~scheme requires much more operations per time step than \DF ~or \mCN. For this numerical application, the CPU time of each approach, using $\Delta t \egal 10^{\,-4}$, has been evaluated using \texttt{Matlab} platform on a computer with Intel i7 CPU and 8GB of RAM and reported in Table~\ref{tab:CPU_time}. It has been preferred to focus on the ratio of computer run time rather than on absolute values,  that are system-dependent. However as expected, the \DF ~scheme is twice faster than the \mCN ~one. The average number of sub-iterations is $\O(N_{\,\mathrm{NL}}) \egal 3$ and as expected, the CPU time of the standard \CN ~scheme is six time longer. In addition, another computational advantage of explicit schemes over the implicit ones is their ease to be parallelised. They allow to achieve almost perfect scaling on high-performance computer systems \cite{Chetverushkin2012}.

Resuming these results, the main advantages of the \DF ~schemes are (i) to avoid sub-iterations to treat the nonlinearities and thus be faster than the \CN ~approach, (ii) to compute \emph{explicitly} the solution at each time step, (iii) to be unconditionally stable, as well as (iv) the ease to be parallelised. 

Uniform discretization approaches are commonly used in building simulation codes such as \textsc{Delphin} \cite{BauklimatikDresden2011} or \textsc{WUFI} \citep{IBP2005} to name a few.  They allow to meet a prescribed tolerance parameter. In the case of a conditionally stable scheme, the time step limitation is imposed by a CFL-type condition to be met. In other cases, the constraint is imposed by the sharp variations of the field, which has been illustrated in Figs.~\ref{fig_AN1:err_fdt} with the choice of the time step according to the variation of the boundary condition. This limitation comes indirectly from the solution accuracy requirement. Comparing it to a scheme with a uniform grid (in space), it might allow to reduce the CPU-time of the code by redistributing the nodes. On the other hand, non-uniform grids might degrade the solution accuracy as illustrated in \cite{Jianchun1995}. Non-uniform grids can be also combined with the \DF ~numerical scheme. However, their implementation requires further development in order to not reduce the order of accuracy, which is certainly an interesting theme for further investigation.

%%% ------------------------------------------------------------------------ %%%

\subsection{Further nonlinear case studies}
\label{sec:case_nonlinear2}

Previous subsection illustrated the relevancy of using the explicit \DF ~scheme to compute the solution of moisture transfer through porous material. The purpose is now to explore the use of this numerical scheme for further case studies, typical cases of moisture transfer in building materials \cite{Janssen2014}. The length of the material is fixed to $L \egal 0.1$ $\mathsf{m}\,$. The initial vapour pressure is $\Pvi \egal 1.16 \cdot 10^{\,3}$ $\mathsf{Pa}\,$, equivalent to a relative humidity $\phi \egal 0.5$. To test the robustness of the scheme, with strong nonlinearities, the properties of a load bearing material are taken from the HAMSTAD benchmark 4 \cite{Hagentoft2004} and are considered for both cases. 
For each one, the solution is compared to a sufficiently converged solution obtained with an \textsc{Euler} explicit scheme. Dimensionless values for each case can also be found in Appendix~\ref{annexe:dimensionless}.

%%% ------------------------------------------------------------------ %%%

\subsubsection{Driving rain case}
\label{sec:rain_case}

The first additional case represents the increase of moisture in the material caused by driving rain at one of the bounding surfaces. For this, at $x \egal 0$ $\mathsf{m}$ a moisture flux $\glL \egal 3.4$ $\mathsf{kg/m^2/s}$ is imposed and there is no transfer with the ambient air. The relative humidity of the ambient air varies according to a sinusoidal variation, with an amplitude of $0.2$, a frequency of $1$h and a mean of $0.5$, at $x \egal 0.1$ $\mathsf{m}$. The  convective vapour coefficient equals $3 \cdot 10^{-8}$  $\mathsf{s/m}$ and the final simulation time is $30$ $\mathsf{h}$.

%%% ------------------------------------------------------------------ %%%

\subsubsection{Capillary adsorption case}
\label{sec:capillary_case}

This case simulates a capillary adsorption of the material. Thus, the left bounding surface of the material is maintained at a saturated state ($\phi \egal 1$). In such case, the \textsc{Robin} type boundary condition Eq.~\eqref{eq:bc} is modified in order to get a \textsc{Dirichlet} one:
\begin{align*}
\Pv &\egal \PvL \,.
\end{align*}
Here, $\PvL \egal 2.33 \cdot 10^{\,3}$ $\mathsf{Pa}$. The vapour pressure of the ambient air is maintained constant at $\PvR \egal 1.16 \cdot 10^{\,3}$ $\mathsf{Pa}$, at $x \egal 0.1$ $\mathsf{m}$, with a convective vapour coefficient set to $3 \cdot 10^{-8}$  $\mathsf{s/m}$. The solution is computed for a final simulation time of $1$ $\mathsf{h}$.

%%% ------------------------------------------------------------------ %%%

\subsubsection{Results and discussion}

\begin{figure}
  \centering
  \subfigure[][\label{fig_AN3:Pv_ft}]{\includegraphics[width=0.48\textwidth]{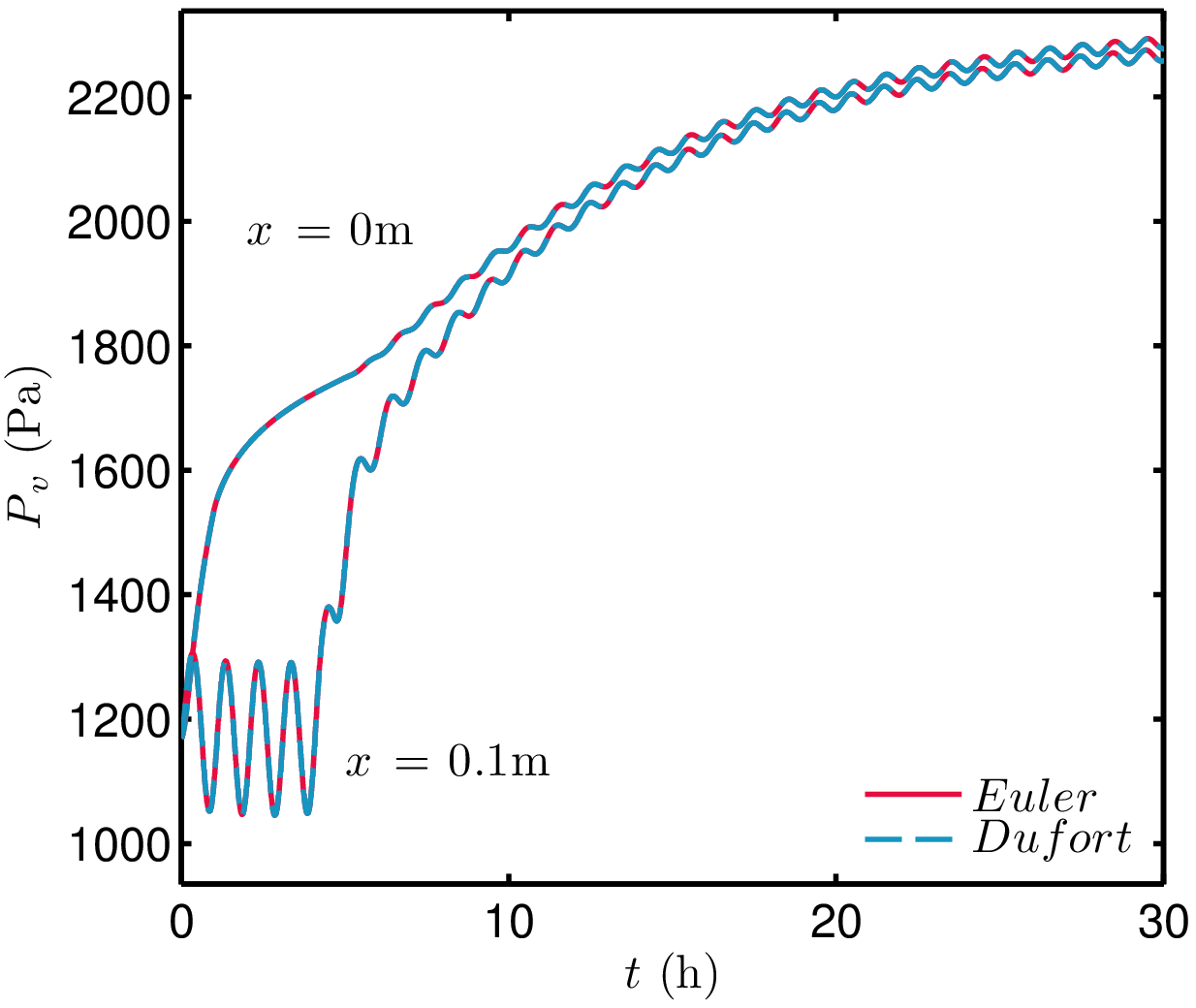}} \hspace{0.3cm}
  \subfigure[][\label{fig_AN4:Pv_fx}]{\includegraphics[width=0.48\textwidth]{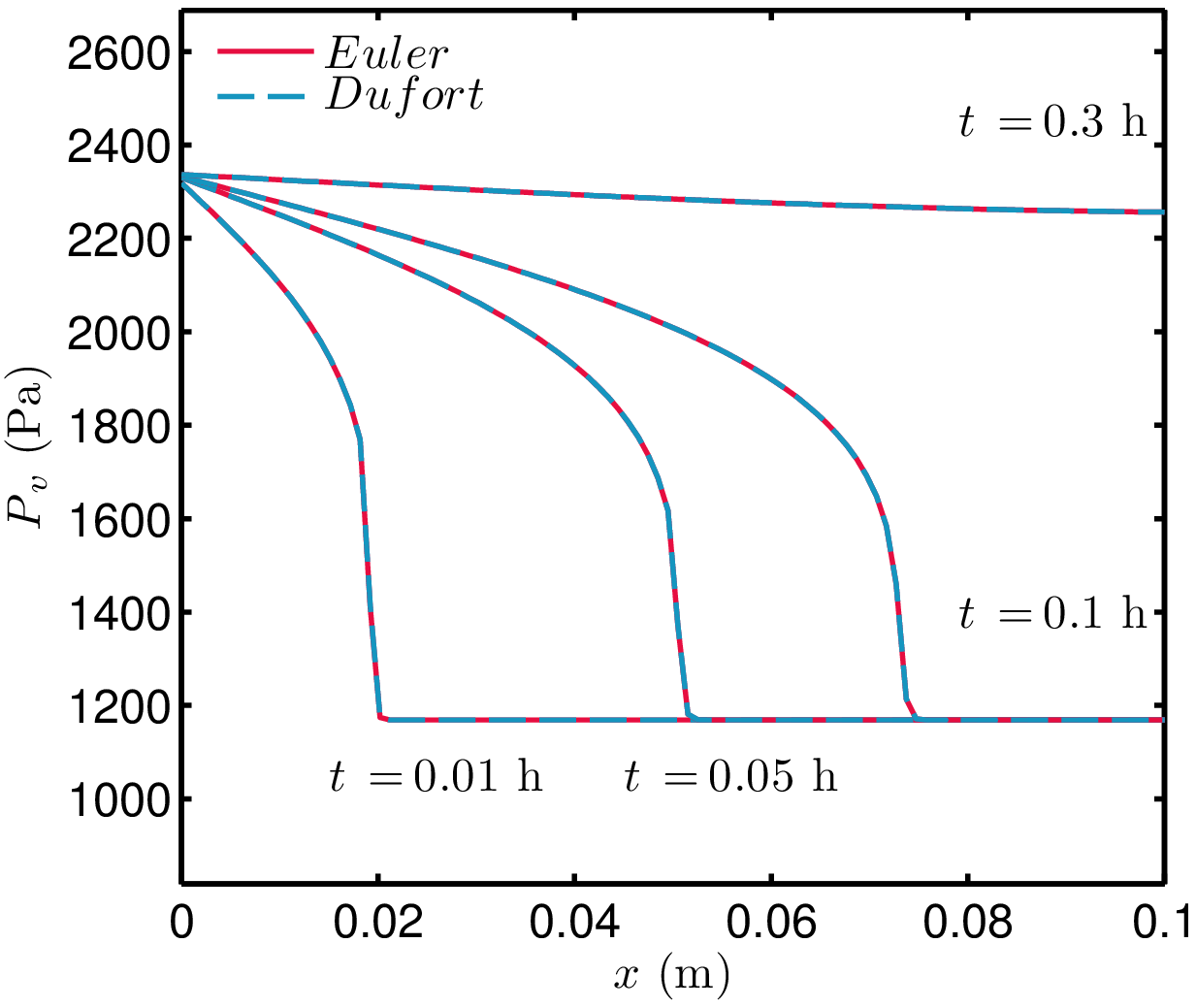}}
  \caption{\small\em Vapour pressure time evolution for the driving rain benchmark (a) and profiles for the capillary adsorption benchmark (b).}
\end{figure}

The discretisations used to compute the solution of both cases are $\Delta x \egal 0.01$ and $\Delta t \egal 10^{-5}$. The evolution of the vapour pressure at $x \egal 0$ $\mathsf{m}$ and $x \egal 0.1$ $\mathsf{m}$ is given in Figure~\ref{fig_AN3:Pv_ft} for the driving rain benchmark. At $x \egal 0$ $\mathsf{m}$, the vapour pressure increases due to the constant rain flux $\glL$ imposed at the surface. At $x \egal 0.1$ $\mathsf{m}$, for $t \leqslant 5$ $\mathsf{h}$, the vapour pressure varies according to the sinusoidal fluctuations of the boundary conditions. Then, as the moisture from the rain flux has diffused through the material, the vapour pressure starts increasing after $t \egal 5$ $\mathsf{h}$. By $t \egal 30$ $\mathsf{h}$, the whole material is saturated. 

\begin{figure}
  \centering
  \includegraphics[width=0.6\textwidth]{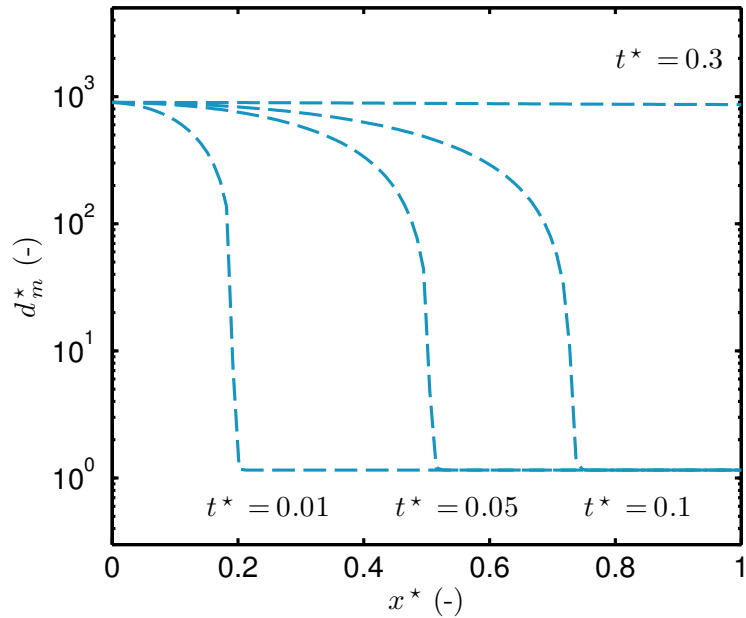}
  \caption{\small\em Profiles of the dimensionless moisture transfer coefficient $\dms$ for the capillary adsorption benchmark.}
  \label{fig_AN4:profil_dMs}
\end{figure}

The vapour pressure profiles at different times are illustrated in Figure~\ref{fig_AN4:Pv_fx} for the capillary adsorption benchmark. At $x \egal 0$ $\mathsf{m}$, the vapour pressure is fixed to the saturation state. The pressure diffuses in the material along the time until reaching the saturation state in the whole material at $t \egal 0.3$ $\mathsf{h}$. The dimensionless moisture transfer coefficient $\dms$, represented in Figure~\ref{fig_AN4:profil_dMs}, highlights the nonlinearities of the material properties. The coefficient has $\O(10^3)$ orders of variation during the simulation. 

\begin{figure}
  \centering
  \includegraphics[width=0.6\textwidth]{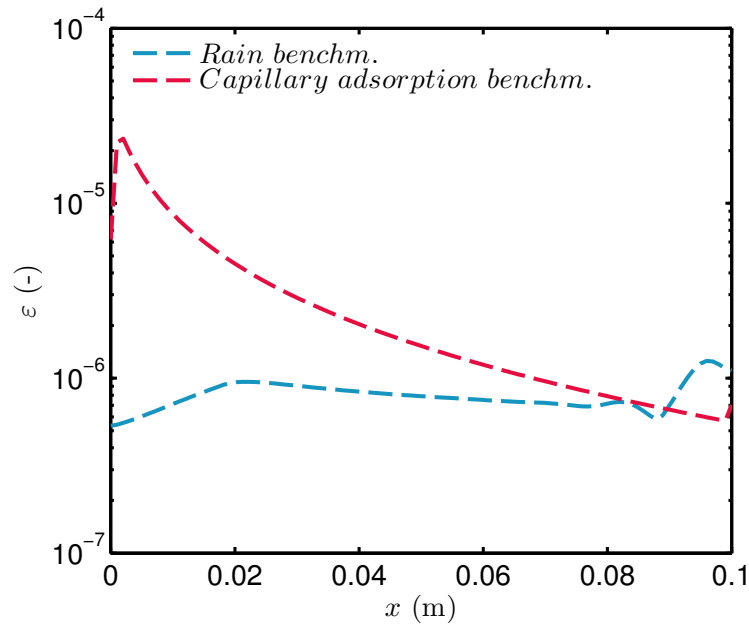}
  \caption{\small\em $\mathcal{L}_{\,2}$ error for the driving rain and the capillary adsorption benchmarks.}
  \label{fig_AN34:err_fx}
\end{figure}

The $\mathcal{L}_{\,2}$ error has been computed between the numerical solution obtained with the \DF ~scheme and the reference one for both cases and illustrated in Figure~\ref{fig_AN34:err_fx}. The error is of the order $\O(10^{-6})$ and proves the high accuracy of the solution computed with the explicit \DF ~scheme. Furthermore, the solution has been computed using the standard \CN ~scheme,  with a tolerance set to $\epsilon \, \leqslant \, 10^{\,-2} \cdot \Delta t^2$, and the CPU times of the schemes are given in Table~\ref{tab:CPU_time}. For both cases, the algorithm using the \CN ~scheme requires around $5$ sub-iterations per time step to meet the tolerance. The \DF ~scheme computes the solution almost ten time faster. These additional results enhance the relevance of using the explicit \DF ~scheme to compute the solution of moisture transfer problems.

\begin{table}
\caption{\small\em Computational time required for the numerical schemes.}
{
\setlength{\extrarowheight}{.3em}
\begin{tabular}[l]{@{}lcc}
\hline
\multicolumn{3}{c}{Hygroscopic adsorption} \\
Numerical Scheme & CPU time ($\mathsf{s}$) & Average number of iterations \\
\DF & 87 & 0 \\
\mCN & 190 & 0 \\
\CN & 550 & 3 \\
\hline
\multicolumn{3}{c}{Driving rain} \\
Numerical Scheme & CPU time ($\mathsf{s}$) & Average number of iterations \\
\DF & 284 & 0 \\
\CN & 2220 & 5 \\
\hline
\multicolumn{3}{c}{Capillary adsorption} \\
Numerical Scheme & CPU time ($\mathsf{s}$) & Average number of iterations \\
\DF & 180 & 0 \\
\CN & 1410 & 5 \\
\hline
\end{tabular}}
\label{tab:CPU_time}
\end{table}

%%% ------------------------------------------------------------------------ %%%

\section{Conclusion}

Most of the Numerical methods applied to mathematical models used in building physics are commonly based on implicit schemes to compute the solution of diffusion problems. The main advantage is due to the stability conditions for the choice of the time discretisation $\Delta t$. However, implicit schemes require important sub-iterations when treating nonlinear problems. This work was therefore devoted to explore some improved schemes and more specifically, the \DF, the \CN ~and the hyperbolisation schemes. The first one is first- or second-order accurate in space, depending on the choice of $\Delta t$ and has the advantage of being unconditionally stable. The second one is also unconditionally stable and second-order accurate in space and time. The latter is second-order accurate in time and in space $\O(\Delta t^2)$ and consistent with the hyperbolic diffusion equation. 

The first case study considered a linear diffusive transfer through a porous material. The \DF, \CN ~and hyperbolisation schemes were compared to the classical \Eu ~explicit scheme and to a reference solution obtained using \textsc{Chebyshev} functions. Results have shown that the hyperbolisation scheme has a stability condition higher than the standard CFL. The error of this scheme depends on parameter $\tau$ representing the amount of hyperbolicity added in the equation. An optimal choice seems to be $\tau \egal \Delta t\,$. As expected, the \DF ~and \CN ~schemes are unconditionally stable and enable to compute the solution for any choice of the time discretisation $\Delta t\,$. In addition, for $\Delta t \leqslant 10^{-3}$, the error of the  \DF ~scheme is first-order accurate in time, and for $\Delta t \geqslant 10^{-3}$ second-order accurate in time. The first conclusions revealed that for $\Delta t \leqslant \dfrac{\Delta x^{\,2}}{2\,\nu}$, it is preferable to use the hyperbolisation scheme, for its accuracy. For larger $\Delta t$ values, or for nonlinear cases, the \DF ~and \CN ~schemes are preferable due to their stability. 

The second case study focused on nonlinear transfer model, with material properties strongly dependent on the vapour pressure field. The extension of the \DF ~and \CN ~schemes were given specially to treat the nonlinearities of the problem. A \mCN ~was proposed in order to avoid sub-iterations at each time step of the algorithm. Both \DF ~and \mCN ~schemes were used to compute the solution of the problem. Results have shown that the error is proportional to $\O(\Delta t)$. The \mCN ~is twice longer than the \DF ~to compute the solution, due to the operations required to compute the implicit and explicit parts of the scheme. The main advantages of the \DF ~schemes is (i) to avoid sub-iterations to treat the nonlinearities, (ii) to compute \emph{explicitly} the solution at each time step, (iii) the unconditionally stable property, as well as (iv) the ease to be parallelised. Additional case-studies with stronger nonlinearities and sharper profiles were analysed, enhancing the advantages of such approach. Attention should be paid for every scheme unconditionally stable because the choice of the time discretisation $\Delta t$ is an important issue to represent accurately the physical phenomena. As mentioned in \cite{Patankar1980}:
\begin{quote}
  \textit{An inexperienced user often interprets this [the unconditionally stable property] to imply that a physically realistic solution will result no matter how large is the time step, and such user is, therefore surprised to encounter oscillatory solutions. The 'stability' in a mathematical sense simply ensures that these oscillations will eventually die out, but it does not guarantee physically plausible solutions.}
\end{quote}

Some examples of unrealistic solutions and some advices on the choice of $\Delta t$, considering the time variations of the boundary conditions, were provided in this study. Keeping this in mind, the \DF ~scheme is a valuable option to compute the solution of nonlinear problems of moisture diffusion in porous materials. Results are encouraging for the use of this approach for treating problems of coupled heat and moisture transfer in two- or three-dimensions and explore other methods such as proposed by \textsc{Saulyev} in \cite{Saulyev1960} to integrate parabolic equations.

%%% ------------------------------------------------------------------------ %%%

\section*{Acknowledgements}

The authors acknowledge the Brazilian Agencies CAPES of the Ministry of Education and CNPQ of the Ministry of Science, Technology and Innovation, for the financial support.

%%% ------------------------------------------------------------------------ %%%

\section*{Nomenclature}

\begin{tabular*}{0.7\textwidth}{@{\extracolsep{\fill}} | c  l l| }
\hline
\multicolumn{3}{|c|}{\emph{Latin letters}} \\
$c_{\,m}$ & moisture storage capacity & $[\mathsf{kg/m^3/Pa}]$ \\
$g$ & liquid flux & $[\mathsf{kg/m^2/s}]$ \\
$h_{\,v}$ & vapour convective transfer coefficient & $[\mathsf{s/m}]$ \\
$k$ & permeability & $[\mathsf{s}]$ \\
$L$ & length & $[\mathsf{m}]$ \\
$\Pc$ & capillary pressure & $[\mathsf{Pa}]$ \\
$\Ps$ & saturation pressure & $[\mathsf{Pa}]$ \\
$\Pv$ & vapour pressure & $[\mathsf{Pa}]$ \\
$R_v$ & water gas constant & $[\mathsf{J/(kg\cdot K)}]$\\
$T$ & temperature & $[\mathsf{K}]$ \\
\multicolumn{3}{|c|}{\emph{Greek letters}} \\
$\phi$ & relative humidity & $[-]$ \\
$\rho$ & specific mass & $[\mathsf{kg/m^3}]$ \\
\hline
\end{tabular*}

%%% ------------------------------------------------------------------------ %%%

\newpage
\appendix
\section{Stability analysis of the \DF ~scheme}
\label{annexe:DF_analysis}

The stability analysis is performed for the Initial Value Problem of the linear diffusion equation:
\begin{align}\label{eq_annx:heat}
& \pd{u}{t} \moins \nu \, \pd{^{\,2} u}{x^{\,2}} \egal 0\,,  && x \ \in \ \mathds{R} \,, && \nu \ > \ 0 \,.
\end{align}
The \DF ~numerical scheme yields to: 
\begin{align}\label{eq_annx:dufort}
& \frac{u_{\,j}^{\,n+1}\ -\ u_{\,j}^{\,n-1}}{2\,\Delta t}\egal \nu\;\frac{u_{\,j-1}^{\,n}\ -\ \bigl(u_{\,j}^{\,n-1}\ +\ u_{\,j}^{\,n+1}\bigr)\ +\ u_{\,j+1}^{\,n}}{\Delta x^{\,2}}\,, \qquad j\ \in \ \mathds{Z} \,, \qquad n\ >\ 0\,.
\end{align}
Using the \textsc{von Neumann}'s stability analysis \cite{Charney1950}, we seek for discrete plane wave solutions of the form:
\begin{align}
u_{\,j}^{\,n} \egal \alpha \, u_{\,0} \, \mathrm{e}^{\,\mathrm{i} \, j \, k \, \Delta x} \,,
\end{align}
where $k \ \Delta x$ is the wave length ($k \, \geqslant \, 1$), $u_{\,0}$ is the initial wave amplitude ($t \egal 0$) and $\alpha$ is the amplification factor between two successive time layers.

In general $\alpha \egal \alpha(\, k \, \Delta x \,)$ and we shall compute this factor for the \DF ~scheme, knowing that:
\begin{align*}
& u_{\,j}^{\,n+1} \egal \alpha \, u_{\,j}^{\,n}  \,,
&& u_{\,j}^{\,n-1} \egal \frac{1}{\alpha} \, u_{\,j}^{\,n} \,,
&& u_{\,j \pm 1}^{\,n} \egal \mathrm{e}^{\,\pm \, \mathrm{i}  \, k \, \Delta x} \, u_{\,j}^{\,n} \,.
\end{align*}
Thus, Eq.~\eqref{eq_annx:dufort} becomes:
\begin{align}
\label{eq_annx:VN_stability}
\biggl(\,\alpha \moins \frac{1}{\alpha} \,\biggr) \egal \mu \, \biggl(\, 2 \, \cos(\sigma) \moins \alpha \moins \frac{1}{\alpha}\,\biggr) \,,
\end{align}
where $\sigma \eqdef k \, \Delta x $ and $\mu \eqdef \frac{2 \, \nu \, \Delta t}{\Delta x^2}$. Eq.~\eqref{eq_annx:VN_stability} admits two roots:
\begin{align*}
& \alpha^{\,+} \egal \frac{\mu \, \cos(\sigma) \plus \sqrt{1 \moins \mu^{\,2} \, \sin^{\,2}(\sigma)}}{1 \plus \mu} \,, \\
& \alpha^{\,-} \egal \frac{\mu \, \cos(\sigma) \moins \sqrt{1 \moins \mu^{\,2} \, \sin^{\,2}(\sigma)}}{1 \plus \mu} \,, \\
\end{align*}

\begin{figure}
\centering
\subfigure[a][\label{fig_annx:alpha_mu1}]{\includegraphics[width=0.48\textwidth]{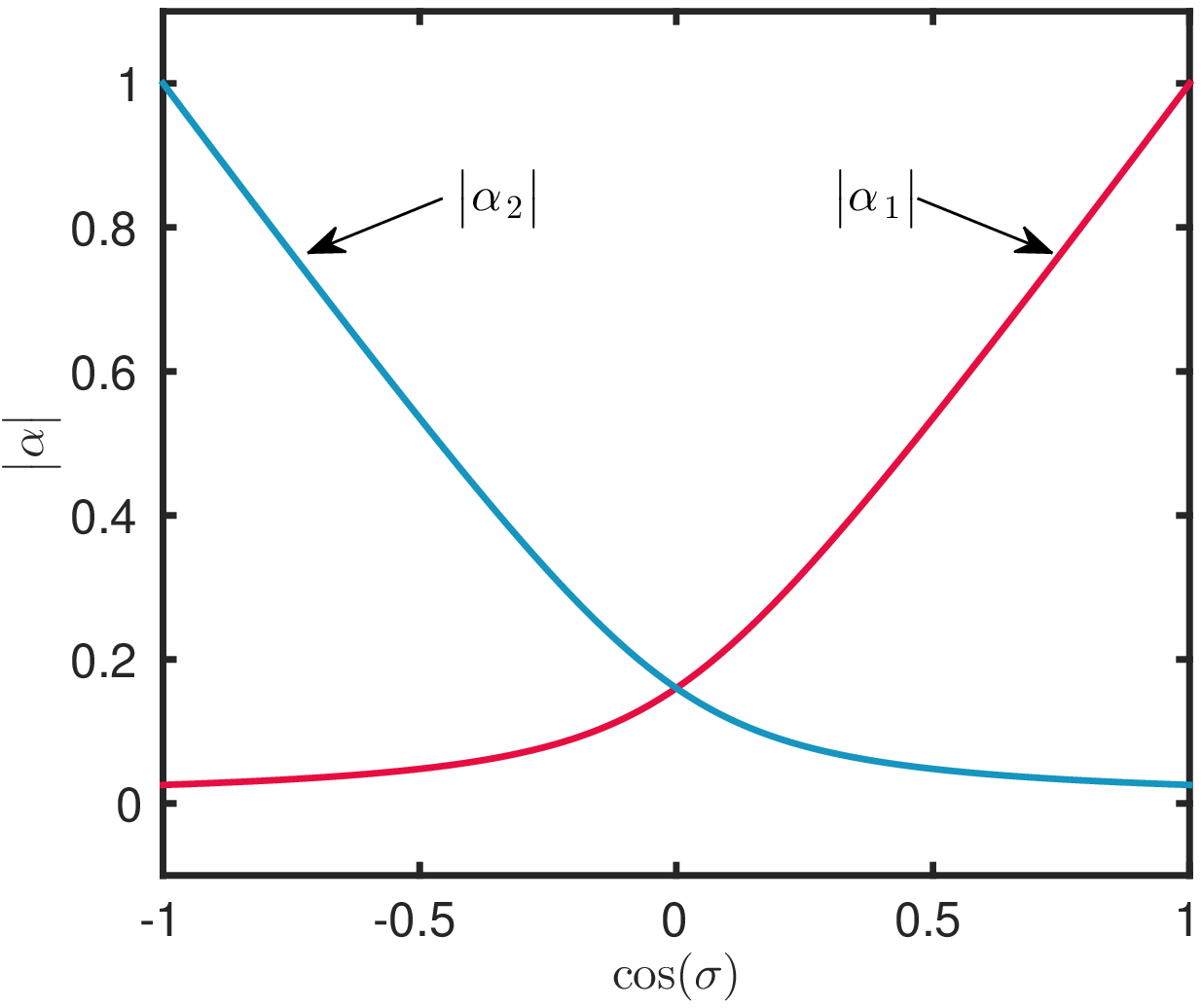}} \hspace{0.3cm}
\subfigure[b][\label{fig_annx:alpha_mu2}]{\includegraphics[width=0.48\textwidth]{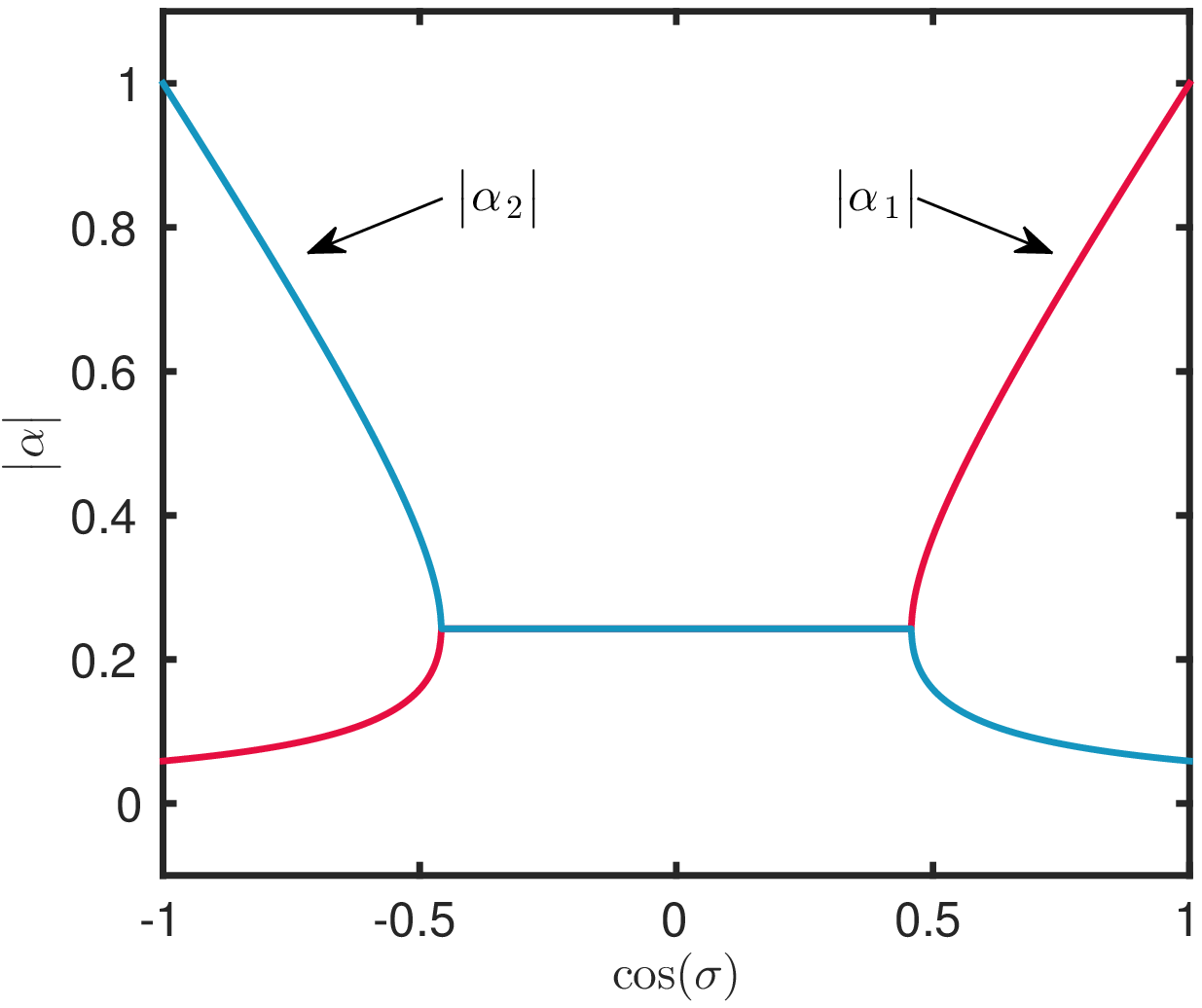}}
\caption{\small\em Amplification factors for the \DF ~scheme for $\mu \egal 0.95$ (a) and  $\mu \egal 1.125$ (b).}
\end{figure}

In Figures~\ref{fig_annx:alpha_mu1} and \ref{fig_annx:alpha_mu2}, we show the dependence of the absolute value of the amplification factors $\bigl|\,\alpha^{\,\pm} \,\bigr|$ on $\cos(\sigma)$ for two fixed values of the parameter $\mu$ --- one below and one above the critical CFL value $\mu\egal 1\,$. It is important to notice that $\bigl|\,\alpha^{\,\pm} \,\bigr| \, \leqslant\, 1$ for all values of the parameter $\sigma\,$. It can be shown (see \cite{Richtmyer1967}) that this property holds for all values of $\mu\,$. This observation completes the proof of the unconditional stability of the DF scheme. The Boundary Value Problem (BVP) for equation Eq.~\eqref{eq_annx:heat} was studied in \cite{Taylor1970}. Namely, it was shown that the same conclusions on the unconditional stability of the \DF ~scheme hold for BVP demonstrated that the boundary conditions are appropriately  discretised.

%%% ------------------------------------------------------------------------ %%%

\section{Dimensionless parameters}
\label{annexe:dimensionless}

\subsection{Case from Section~\ref{sec:case_linear}}

Problem~\eqref{eq:moisture_dimensionlesspb_1D} is considered with $\glsL \egal \glsR \egal 0\,$. The dimensionless properties of the material are equal to $\dms \egal 1$ and $\cms \egal 8.6\,$. The final simulation time is fixed to $\tau^{\,\star} \egal 120$, while Biot numbers are $\BivL \egal 101.5$ and $\BivR \egal 15.2\,$. The boundary conditions are expressed as:
\begin{align*}
& \uL (\ts) \egal 1 \plus \frac{1}{2} \; \sin \left(\, \frac{2\pi \, \ts}{24}\,\right) 
\plus \frac{1}{2} \; \sin \left(\, \frac{2\pi \, \ts}{4}\,\right) \,,\\
& \uR (\ts) \egal 1 \plus \frac{4}{5} \;\sin \left(\, \frac{2\pi \, \ts}{12}\,\right) \,.
\end{align*}

\subsection{Case from Section~\ref{sec:case_nonlinear1}}

The dimensionless properties of the materials are:
\begin{align*}
& \dms \egal 1 \plus 0.91 \, u \plus 600 \cdot \exp \Bigl[ \,-10 \, \bigl(\, u \moins 1.9 \, \bigr)^2 \, \Bigr] \,,\\
& \cms \egal 900 \moins 656 \, u \plus 10^4 \cdot \exp \Bigl[ \,-5 \, \bigl(\, u \moins 1.9 \, \bigr)^2 \,  \Bigr] \,.
\end{align*}
The Biot numbers are also assumed to be $\BivL \egal 101.5$ and $\BivR \egal 15.2 \,$. The ambient water vapour pressure at the boundaries are different from the previous case study. At the left boundary, $\uL$ has a fast drop until the saturation state $\uL \egal 2, \; \forall t \in  \bigr[ \, 10 , \, 40 \, \bigl] $ and at the right boundary, $\uR \egal 1 \plus 0.8 \; \sin \left(\, 2\pi \, \ts \,\right) \,.$

\subsection{Case from Section~\ref{sec:case_nonlinear2}}

For both case-studies, an analytic expression of the material properties has been fitted, which dimensionless formulation is:
\begin{align*}
\dms & \egal 0.85 \, u^{\, -0.71} + 900 \,  \exp\Bigl[ \,-8 \, \bigl(\, u \moins 2 \, \bigr)^2 \, \Bigr] \,, \\
\cms & \egal 1.69 \cdot 10^{\,2} \, u^{\, -0.53} + 3 \exp\Bigl[ \,-9 \, \bigl(\, u \moins 1.3 \, \bigr)^2 \, \Bigr] \,.
\end{align*}
\
The dimensionless parameters for the driving rain case (Section~\ref{sec:rain_case}) are: 
\begin{align*}
& \glsL \egal 14.7 \,,
&& \BivL \egal 0 \,,
&& \BivR \egal 15.2 \,,
&& \uR \egal 1 + 0.4 \sin \bigl(\, 2 \ \pi \ \ts \,\bigr) \,,
&& \tau^{\,\star} \egal 30 \,,
\end{align*}
\
while, for the capillary adsorption case (Section~\ref{sec:capillary_case}), they are:
\begin{align*}
& uL \egal 2 \,,
&& \BivR \egal 15.2 \,,
&& \uR \egal 1 \,,
&& \tau^{\,\star} \egal 1 \,.
\end{align*}

%%% ------------------------------------------------------------------------ %%%

%%% Bibliography
\addcontentsline{toc}{section}{References}
\bibliographystyle{abbrv}
\bibliography{biblio}

\end{document}